\documentclass[reprint, superscriptaddress, amsmath,amssymb, aps, pre]{revtex4-2}
\usepackage[T1]{fontenc}
\usepackage{graphicx}
\usepackage{dcolumn}
\usepackage{bm}
\usepackage{hyperref}
\usepackage{xcolor}

\begin{document}

\title{Inertial flows over the Koch leaf}

\author{Sahrish B. Naqvi}
\email{sahrishbatool.naqvi@uwr.edu.pl}
\affiliation{Institute of Theoretical Physics, Faculty of Physics and Astronomy, University of Wroc{\l}aw, pl. M. Borna 9, 50-204 Wroc{\l}aw, Poland}
\author{Dawid Strzelczyk}
\affiliation{Institute of Theoretical Physics, Faculty of Physics and Astronomy, University of Wroc{\l}aw, pl. M. Borna 9, 50-204 Wroc{\l}aw, Poland}
\author{Maciej Tadej}
\affiliation{Mathematical Institute, University of Wroc{\l}aw, pl. Grunwaldzki 2, 50-384 Wroc{\l}aw, Poland}
\author{Maciej Matyka}
\email{maciej.matyka@uwr.edu.pl}
\altaffiliation[Also at ]{Jo\v{z}ef Stefan Institute, Jamova cesta 39, 1000 Ljubljana, Slovenia}
\affiliation{Institute of Theoretical Physics, Faculty of Physics and Astronomy, University of Wroc{\l}aw, pl. M. Borna 9, 50-204 Wroc{\l}aw, Poland}

\date{\today}

\begin{abstract}

The main hypothesis of this paper is that inertial effects known from porous-media flow can be observed even in a simple system consisting of a single channel. The strength of inertia is controlled by the Reynolds number, $Re$. We investigated fluid flow through a channel with a cavity in the shape of a Koch leaf. The Koch leaf of level-6 was chosen because it combines regularity, repeatability, and representativeness with a complex multiscale structure and non-trivial hydrodynamic interactions. We analyzed nonlinear effects through the so-called friction factor and show that its deviation from Darcy's law is very subtle. We further examined the influence of inertia on the flow structure, including the participation number $\pi$ and the backward-flow fraction $\rho^{-}$. In contrast to disordered porous media, the participation number decreases with $Re$, indicating that inertia concentrates kinetic energy in the main through-flow, while recirculation (backward flow) increasingly occupies the Koch leaf cavity. The tortuosity varies non-monotonically with $Re$, the apparent permeability decreases, and in the inertial regime the two are related by a power law with an exponent comparable to that found for natural rock. At the lowest and highest Reynolds numbers, the local kinetic energy followed a power-law distribution, $P(\varepsilon)\propto\varepsilon^{-m}$, over several orders of magnitude with the same exponent, $m\approx0.83$; inertia modified only the tails of the distribution. Our research can be regarded as an investigation of transport in an extremely porous system with porosity $\varphi \approx 1$.
\end{abstract}

\maketitle


\section{\label{sec:intro}INTRODUCTION}

Fluids flowing through porous media self-organize into preferential flow channels, whose structure is set by the interaction between the moving fluid and the geometry of the fluid--solid interface \cite{hyman2020flow,andrade1999inertial}. When the driving force is large, the flow along these channels may gain enough momentum for inertial effects to appear, even though the flow remains laminar. Understanding how such channels form and how they are modified by inertia is important for subsurface energy storage and CO$_2$ sequestration \cite{hovorka2019dynamic}, groundwater abstraction \cite{matyka2017granica}, and engineered porous systems such as microfluidic devices \cite{cao2019application}, porous-medium combustors \cite{banerjee2021developments}, filters \cite{zhang2024optimal}, and catalytic and thermochemical reactors \cite{wang2014thermal}.

At low flow velocities, fluid transport through a porous medium is described by Darcy's law, which establishes a linear relationship between the applied pressure gradient and the flow velocity \cite{neuman1977theoretical,whitaker1986flow},
\begin{equation}
-\nabla p = \frac{\mu}{k}\,\mathbf{U},
\label{eq:darcy-law}
\end{equation}
where $k$ is the permeability and $\mathbf{U}$ is the superficial velocity. As the velocity increases, inertial effects become significant and deviations from linearity emerge. These are commonly captured by the Forchheimer equation, which introduces a second-order correction \cite{whitaker1996Forchheimer,barree2004beyond},
\begin{equation}
-\nabla p = \frac{\mu}{k}\,\mathbf{U} + \beta\rho\,|\mathbf{U}|\,\mathbf{U},
\label{eq:forchheimer}
\end{equation}
where $\beta$ is the non-Darcy (inertial) coefficient. The transition between the two regimes is governed by the Reynolds number and involves a number of pore-scale phenomena, including local flow separation, vortex formation, and enhanced drag \cite{fourar2005inertia,arthur2020piv,balhoff2009predictive}. Its onset and strength depend sensitively on the properties of the medium, in particular on porosity and heterogeneity \cite{zeng2006criterion,arbabi2024transition,sniezek2024inertia}. Pore-scale simulations of bead packs and natural rocks have shown that heterogeneous samples can depart from Darcy flow at Reynolds numbers two to three orders of magnitude lower than homogeneous ones of similar porosity \cite{muljadi2016impact}, and heterogeneity also modifies how inertia affects solute transport \cite{nissan2018inertial}. Deviations from Darcy's law have even been reported at very low Reynolds numbers, in the so-called pre-Darcy regime \cite{basak1977non,wang2019experimental}.

A further difficulty is that the onset of inertial effects does not necessarily coincide with a pronounced change in the global pressure--flow relation. Changes in the internal organization of the flow can appear considerably earlier and may remain hidden if only integral quantities, such as the friction factor or the apparent permeability, are considered \cite{sniezek2024inertia,Naqvi26}. Pore-scale measures are therefore needed to resolve the transition. The participation number $\pi$ quantifies how kinetic energy is distributed over the pore space \cite{andrade1999inertial}; the hydraulic tortuosity $\tau$ measures how strongly the flow paths deviate from the mean flow direction \cite{Koponen97,Duda11}; and the backward-flow fraction and the formation of vortices characterize recirculation \cite{sniezek2024inertia,Naqvi26}. In disordered media, the transition toward inertia-influenced flow is accompanied by an increase in $\pi$, as kinetic energy is redistributed from preferential channels into the surrounding pore space \cite{andrade1999inertial,sniezek2024inertia}, and by a non-monotonic change in tortuosity \cite{sniezek2024inertia,Naqvi26}.

The statistics of the local kinetic energy provide complementary information. Andrade \textit{et al.} \cite{andrade1997fluid} solved the Navier--Stokes equations in a two-dimensional percolation cluster and found that the distribution of the local kinetic energy, $n(E)\propto E^{-\alpha}$, follows a power law over roughly five orders of magnitude, with $\alpha = 0.90 \pm 0.03$ for the full cluster and $\alpha = 0.64 \pm 0.05$ for the flow-carrying backbone alone. This behavior contrasts with the log-normal distribution of local currents in the analogous random resistor network and shows that slowly moving fluid in the ``stagnant'' zones plays a significant role in transport through porous media. The exponent remained practically constant over two decades of low Reynolds numbers, but decreased to $\alpha = 0.79 \pm 0.03$ at high $Re$, where several vortices and zones of flow separation appeared \cite{andrade1997fluid}. The energy distribution therefore links the geometry of the slow, near-wall fluid to the onset of inertia, but it has so far been studied mainly in disordered structures.

Most of these observations come from idealized granular packings or from disordered and natural porous media, in which several geometrical properties---porosity, pore-size distribution, connectivity, tortuosity, and surface roughness---act simultaneously. Consequently, it is difficult to distinguish changes produced directly by fluid inertia from those imposed by structural disorder. In a recent study of low-porosity natural rock samples, we showed that the transition to inertial flow occurs through a distinct mechanism, visible as a change in tortuosity and the early formation of vortical structures, while the friction factor shows no clear departure from Darcy's law \cite{Naqvi26}. Rough and self-affine boundaries are likewise known to influence hydraulic transport and permeability \cite{madadi2003rough,madadi2003fractures}. These results raise the question of which features of inertial porous-media flow originate from disorder and which can emerge solely from the nonlinear interaction between fluid momentum and a complex solid boundary.

Deterministic fractal geometries provide a way to address this question in a controlled setting, because they contain geometrical features over several length scales while remaining exactly reproducible. The Koch leaf of any level is a particularly simple example of such a multiscale boundary: at every construction level, smaller geometrical features are introduced according to the same deterministic rule. Higher-order Koch leaf geometries (order $n \geq 3$) therefore represent an intermediate level of geometric complexity between simple obstructions and the highly irregular boundaries encountered in natural porous materials, and have already been investigated in transport-related studies. For instance, Adrover showed that convection-dominated dispersion in channels bounded by Koch leaf and Koch snowflakes exhibits a scaling different from that of smooth boundaries \cite{adrover2011koch}. However, comparatively little attention has been devoted to how an increasing contribution of fluid inertia progressively alters the flow structure around an isolated Koch-type obstruction.

In this work, we investigate steady incompressible flow through a two-dimensional channel containing Koch-leaf of order 6 geometry. In contrast to the low-porosity rocks studied in Ref.~\cite{Naqvi26}, the present system represents the opposite limit: it contains a single deterministic obstruction and has a very high fluid-volume fraction, $\varphi\approx1$. The Reynolds number is systematically increased while the geometry remains unchanged. We examine whether signatures characteristic of inertial porous-media flows emerge in this simplified system by analyzing the macroscopic friction response together with the tortuosity $\tau$, the participation number $\pi$, the backward-flow fraction $\rho^{-}$, the apparent permeability, and the probability distribution of the local kinetic energy. The simulations are complemented by a semi-analytical solution of the Stokes problem, built from Moffatt corner modes of the self-similar Koch geometry, which provides an independent reference for the viscous limit. Our central hypothesis is that measurable inertia-driven changes in the flow structure can develop even in a highly open and geometrically deterministic system, while their signature in conventional macroscopic measures remains comparatively subtle.

\section{\label{sec:materials}MATERIALS AND METHODS}
To investigate the emergence of inertial effects in a geometrically well-defined system, we considered a two-dimensional flow domain incorporating a Koch leaf-shaped boundary connected to the channel walls. The height of the Koch leaf was fixed at $L=0.3\,\mathrm{m}$ and was used as the characteristic length scale throughout the analysis.

\begin{figure}
    \centering
    \includegraphics[width=1\linewidth]{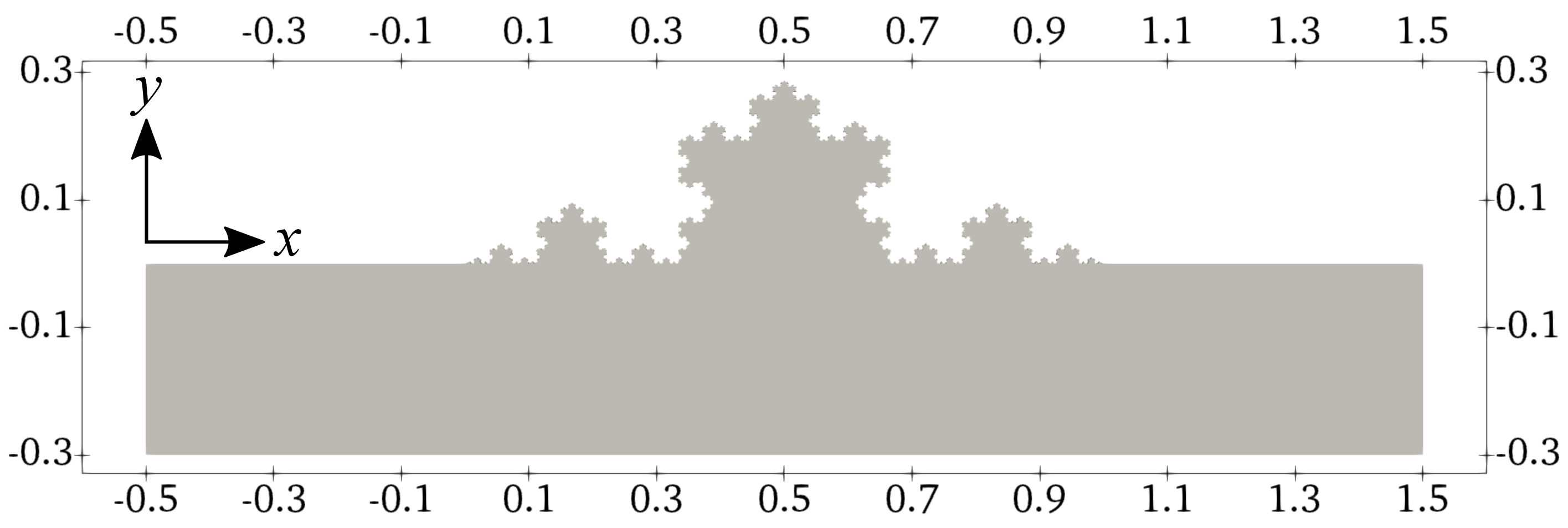}
    \caption{Geometry of the Koch leaf of order 6, where the porosity of the channel together with Koch leaf obstruction is $\varphi \approx 1$.}
    \label{fig:placeholder}
\end{figure}

The fluid was assumed to be Newtonian and incompressible, and the steady-state flow was described by the incompressible Navier--Stokes equations,
\begin{equation}
\nabla \cdot \mathbf{u}=0,
\label{eq:continuity}
\end{equation}
and
\begin{equation}
\rho(\mathbf{u}\cdot\nabla)\mathbf{u}
=
-\nabla p+\mu\nabla^2\mathbf{u},
\label{eq:nse}
\end{equation}
where $\mathbf{u}$ is the velocity field, $p$ is the pressure, $\rho$ is the fluid density, and $\mu$ is the dynamic viscosity.

The flow was imposed at the inlet by prescribing a fully developed parabolic velocity profile. The inlet extended from $y=-0.30\,\mathrm{m}$ to $y=0$, with its centerline located at $y=-0.15\,\mathrm{m}$. The streamwise component of the inlet velocity was prescribed as
\begin{equation}
u_x(y)
=
U_{\max}
\left[
1-
\left(
\frac{y+0.15}{0.15}
\right)^2
\right],
\label{eq:inlet}
\end{equation}
with
\begin{equation}
u_y=0.
\end{equation}
Here, $U_{\max}$ denotes the maximum inlet velocity, attained at the centerline of the inlet. The velocity decreases parabolically toward the inlet edges and becomes zero at $y=-0.30\,\mathrm{m}$ and $y=0$. The value of $U_{\max}$ was varied between simulations to obtain different flow conditions and thereby systematically increase the relative contribution of inertial effects.

No-slip boundary conditions,
\begin{equation}
\mathbf{u}=0,
\end{equation}
were imposed on the Koch surface and all other solid boundaries of the computational domain. At these walls and at the inlet, a zero normal gradient was imposed for the pressure, $\partial p/\partial n=0$. At the outlet, a zero normal gradient was imposed for the velocity, and the pressure was fixed to a reference value,
\begin{equation}
\frac{\partial \mathbf{u}}{\partial n}=0,
\qquad
p=0,
\label{eq:outletBC}
\end{equation}
where $n$ denotes the outward normal direction.
The governing equations were solved using the finite-volume method implemented in OpenFOAM \cite{jasak2007openfoam}. Steady-state solutions were obtained using the \texttt{simpleFoam} solver, which employs the Semi-Implicit Method for Pressure-Linked Equations (SIMPLE) algorithm for pressure--velocity coupling \cite{patankar1980numerical}. Numerical convergence was assessed by monitoring the residuals of the pressure and velocity equations. The calculations were continued until all residuals decreased below $10^{-6}$.

The strength of inertial effects was characterized by the Reynolds number,
\begin{equation}
Re=\frac{U L}{\nu},
\label{eq:Re}
\end{equation}
where $L=0.3\,\mathrm{m}$ is the height of the Koch leaf, $U$ is the mean inlet velocity, and $\nu=\mu/\rho$ is the kinematic viscosity. For the prescribed parabolic inlet profile, the mean inlet velocity is related to the maximum velocity by
\begin{equation}
U=\frac{2}{3}U_{\max}.
\label{eq:umean}
\end{equation}
Consequently, the Reynolds number can be written directly in terms of the imposed maximum inlet velocity as
\begin{equation}
Re=
\frac{2U_{\max}L}{3\nu}.
\label{eq:ReUmax}
\end{equation}
The Koch geometry, characteristic length, and fluid properties were kept unchanged throughout the simulations, while $U_{\max}$ was systematically varied to cover the investigated Reynolds-number range.

The computational geometry was discretized using the \texttt{cartesian2DMesh} utility of the cfMesh library \cite{cfmesh2015}, and the resulting mesh was imported into OpenFOAM v. 2006. Local mesh refinement was applied in the vicinity of the Koch boundary in order to accurately resolve the velocity gradients close to the solid surface.

To obtain a sufficient number of sampling points to generate the probability density functions of the kinetic energy in the low-velocity areas, we oversampled the OpenFOAM solutions using RBF-FD interpolation~\cite{Strzelczyk2025}. In particular, the mesh elements with more than 3 sides were first divided into triangles. Subdivision of the resulting triangles into smaller ones was decided based on the magnitude of the kinetic energy in the triangle's centroid -- regions of lower kinetic energy were subdivided into more triangles. The sampling points were then placed at the centroids of the subdivided triangles. We used cubic radial basis functions augmented with second order monomials as the interpolation basis.


To provide an independent reference for the viscous limit, we also constructed a semi-analytical solution of the flow for $Re\to0$. In this limit, the inertial term in Eq.~\eqref{eq:nse} vanishes and the problem reduces to the Stokes equations,
\begin{equation}
\nabla\cdot\mathbf{u}=0, \qquad \mu\nabla^{2}\mathbf{u}=\nabla p,
\label{eq:stokes}
\end{equation}
with the boundary conditions of Eqs.~\eqref{eq:inlet}--\eqref{eq:outletBC}. Solutions of the Stokes problem have a stream function of the Goursat form $\psi=\mathrm{Re}\{\bar{z}f(z)+g(z)\}$, with $z=x+iy$ and $f$, $g$ analytic \cite{goursat1898}, from which the velocity, vorticity, and pressure follow. Near a corner of opening angle $2\alpha$, the admissible functions are powers $z^{\lambda}$, and the no-slip condition fixes $\lambda$ through
\begin{equation}
\sin\!\left[2(\lambda-1)\alpha\right] \pm (\lambda-1)\sin 2\alpha = 0
\label{eq:moffatt}
\end{equation}
\cite{moffatt1964viscous}. For the $60^\circ$ apices of the Koch triangles the exponents are complex ($\lambda_1\approx5.06+1.95i$), which gives the infinite sequence of Moffatt eddies, whereas the $240^\circ$ corners at the ends of the openings give real exponents ($\lambda=1.616,\,2.149,\ldots$).

In each Koch triangle $T_{k,l}$, the velocity is written as
\begin{equation}
\mathbf{u}=\sum_{n} a_{k,l,n}\,\mathbf{u}^{60^\circ}_{k,l,n}
+\sum_{n} b_{k,l,n}\,\mathbf{u}^{240^\circ}_{k,l,n}
+\sum_{n} c_{k,l,n}\,\mathbf{u}^{\mathrm{reg}}_{k,l,n},
\label{eq:koch-expansion}
\end{equation}
with a separate set of coefficients in every triangle. Here, $\mathbf{u}^{60^\circ}_{k,l,n}$ are the Moffatt modes of the apex of the triangle and $\mathbf{u}^{240^\circ}_{k,l,n}$ the modes of the $240^\circ$ corners at the ends of its opening and of the openings on its sides; each of them satisfies the no-slip condition on the two walls of its own corner. The last sum is the regular part, of Goursat form with $f$ and $g$ polynomial in $z-z_0$, where $z_0$ is the centroid of the triangle. Polynomial Goursat functions are complete: every biharmonic function on a simply connected domain can be approximated by them \cite{vekua1967}. Polynomials alone, however, converge slowly near the corners, where $f$ and $g$ are singular; the corner modes remove these singularities, in the same way as in the lightning Stokes solver \cite{brubeck2022lightning} and in the method of particular solutions \cite{fox1967approximations}. Unlike in Ref.~\cite{brubeck2022lightning}, the corner terms used here are the exact Moffatt modes of each triangle of the self-similar geometry.

In the channel, the solution is decomposed as
\begin{equation}
\mathbf{u}=\mathbf{u}_P+\mathbf{u}_S,
\label{eq:channel-split}
\end{equation}
where $\mathbf{u}_P$ is the Poiseuille profile of Eq.~\eqref{eq:inlet} extended to the whole channel. It satisfies the Stokes equations, the no-slip condition on both channel walls, and the inlet and outlet conditions, and it vanishes on the openings to the cavity. The correction $\mathbf{u}_S$ is the Stokes flow that equals the cavity velocity of Eq.~\eqref{eq:koch-expansion} on the openings, satisfies the outlet condition, and vanishes on the remaining boundaries. Adding this correction preserves all boundary conditions of $\mathbf{u}_P$ and makes the velocity continuous across the openings. Since every term in Eqs.~\eqref{eq:koch-expansion} and \eqref{eq:channel-split} solves the Stokes problem, the coefficients are determined by a least-squares fit to the no-slip condition on the walls and to the continuity of vorticity and pressure across the openings. The solution was computed for the same sixth-generation Koch leaf and the same $2\times0.3\,\mathrm{m}$ channel as the simulations. From it, we evaluated $\tau$, $\pi$, $\rho^{-}$, and $\mathcal{K}_{\mathrm{app}}$ with the same definitions as for the simulated fields [Eqs.~\eqref{eq:tau}, \eqref{eq:participation}, \eqref{eq:rho-minus}, and \eqref{eq:kapp} in Sec.~\ref{sec:results}]; these Stokes-limit values are shown as solid horizontal lines in Figs.~\ref{fig:tau}--\ref{fig:kappa}.

\section{\label{sec:results}RESULTS AND DISCUSSION}
\begin{figure}
    \centering
    \includegraphics[width=1\linewidth]{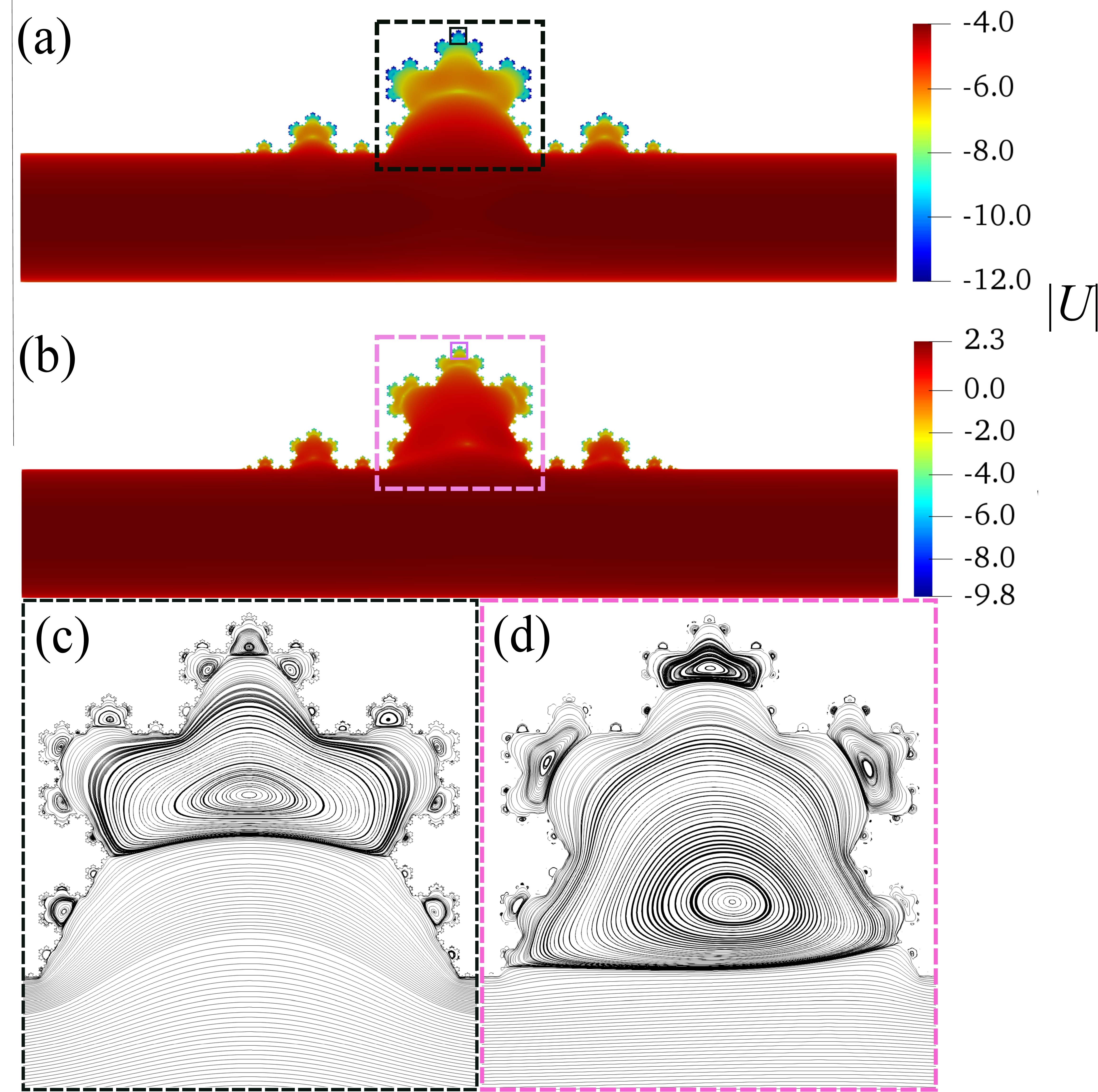}
    \caption{Velocity-magnitude $|U|[m/s]$ fields at the lowest (a) $Re=6.67\times 10^{-5}$ and highest (b) $Re\approx85.09$ Reynolds numbers considered in the simulations. The dashed boxes (c) and (d) indicate the regions enlarged in the lower panels, where the corresponding streamlines are shown. The logarithmic color scale highlights low-velocity regions and illustrates the increasing spatial redistribution of the flow with increasing inertia.}
    \label{fig:velocity-field}
\end{figure}

The velocity-magnitude fields at two representative Reynolds numbers are shown in Fig.~\ref{fig:velocity-field}. At the lowest Reynolds number, $Re=6.67\times10^{-5}$ [Fig.~\ref{fig:velocity-field}(a)], the flow is strongly dominated by viscous effects, and the velocity distribution closely follows the multiscale geometry of the Koch boundary. Broad regions of reduced velocity develop in the vicinity of the solid surface, while the velocity varies smoothly away from the obstruction. As the Reynolds number increases to $Re\approx85.09$ [Fig.~\ref{fig:velocity-field}(b)], the influence of fluid inertia becomes more pronounced. The higher fluid momentum leads to a stronger localization of the velocity gradients near the Koch surface and to a redistribution of the flow around its geometrical features. In particular, the low-velocity regions become more confined to the vicinity of the boundary, while the surrounding flow retains a larger momentum. The comparison therefore indicates a redistribution of the flow with increasing $Re$, motivating a more quantitative analysis of the inertial effects in the following results.

A closer view of the streamlines, shown in the lower panels of Fig.~\ref{fig:velocity-field}, reveals a clear change in the internal flow structure with increasing Reynolds number. At the lowest Reynolds number, several small recirculation pockets appear in the vicinity of the finer geometrical features. At $Re\approx85.09$, the increased fluid momentum reduces the ability of the flow to follow each local variation of the boundary. Consequently, the streamlines in the main channel become straighter, reflecting the increasing importance of advective inertia relative to viscous diffusion. At the same time, the flow inside the Koch cavity changes qualitatively. The main stream separates from the boundary at the entrance of the cavity, and the small, isolated recirculation pockets observed at low $Re$ merge into a single large primary vortex that occupies most of the cavity. This primary vortex is accompanied by smaller secondary vortices trapped within the finer recesses of the Koch boundary. The comparison thus reveals two simultaneous effects of increasing inertia: a straightening of the main through-flow due to momentum preservation and an enhancement of separated recirculating flow inside the Koch cavity. Thus, although the external streamlines become less sensitive to the small-scale boundary geometry, the internal flow structure becomes increasingly dominated by inertia-induced vortical motion.

\begin{figure}
    \centering
    \includegraphics[width=1\linewidth]{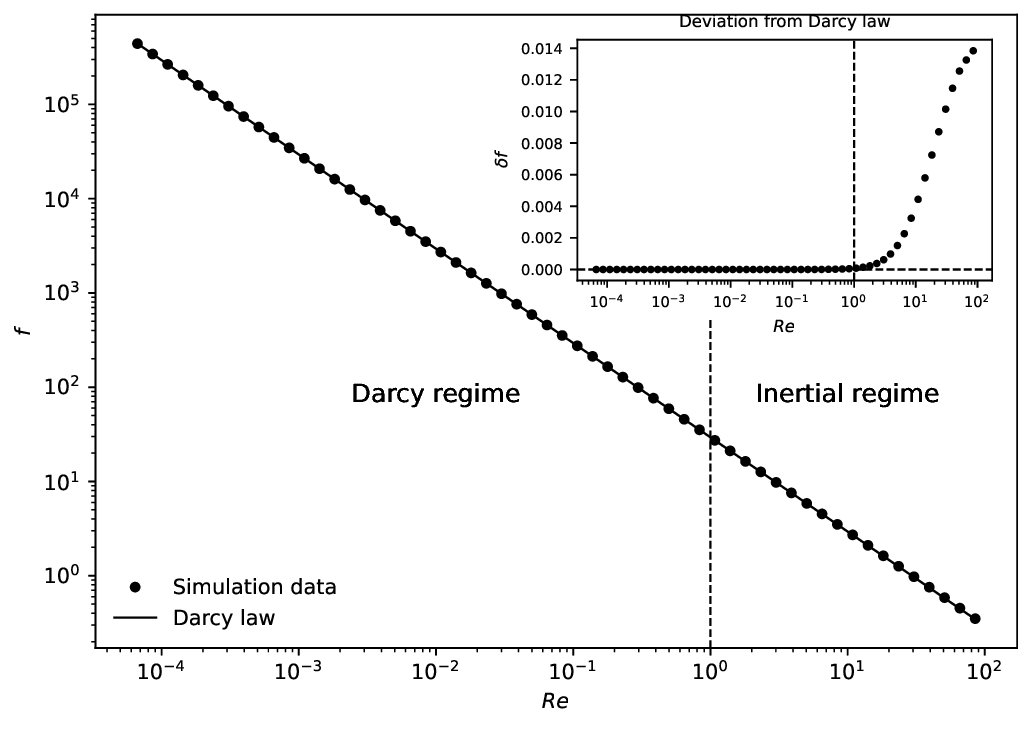}
    \caption{Friction factor versus $Re$. The dashed horizontal line in the inset at $\delta f=0$ indicates no deviation from the Darcy prediction. The vertical dashed line in the main plot and inset serves as a guide, separating the Darcy regime from the inertial regime.}
    \label{fig:FF}
\end{figure}
To assess whether the changes observed in the internal flow structure
are also reflected in the macroscopic flow resistance, we calculated
the friction factor~\cite{Naqvi26} as

\begin{equation}
f=
\frac{\Delta p\,L}
{\langle |\mathbf{u}| \rangle_\Omega^2},
\label{eq:friction}
\end{equation}

where $\Delta p$ is the pressure drop across the computational domain, and $\langle |\mathbf{u}| \rangle_\Omega$ is the area-weighted mean
velocity magnitude,

\begin{equation}
\langle |\mathbf{u}| \rangle_\Omega
=
\frac{\sum_{i=1}^{N} |\mathbf{u}_i|A_i}
{\sum_{i=1}^{N} A_i}.
\end{equation}

Here, $\Omega$ denotes the entire two-dimensional fluid domain, comprising the main channel and the Koch-leaf cavity. The quantity $N$ is the total number of computational cells within $\Omega$, and $A_i$ is the area of the $i$-th cell.
In the viscous-dominated regime, the friction factor follows the Darcy
relation

\begin{equation}
f_D=\frac{b}{Re},
\label{eq:darcy-fit}
\end{equation}

where $b$ is obtained from a fit to the low-$Re$ data. As shown in
Fig.~\ref{fig:FF}, the simulated friction factor remains very close
to the Darcy fit over most of the investigated Reynolds-number
range. Therefore, the influence of inertia is not immediately evident
from the friction-factor curve alone, similar to the behavior reported
for the geological porous medium \cite{Naqvi26}.

To resolve the small departure from the Darcy relation more clearly,
we calculated the relative deviation of the simulated friction factor
from the Darcy prediction as

\begin{equation}
\delta f =
\frac{f-f_D}{f_D},
\label{eq:darcy-deviation}
\end{equation}

where $f$ is the friction factor obtained from the
simulation and $f_D=b/Re$ is the value predicted by the Darcy fit.
At low $Re$, the relative deviation remains close to zero, confirming
that the flow follows the Darcy scaling. With increasing $Re$, the
deviation gradually increases and becomes more pronounced at higher
Reynolds numbers, reaching $\delta f\approx0.014$ (i.e., $1.4\%$) at the largest
$Re$ considered, which indicates the growing contribution of inertial
effects. Nevertheless, the deviation remains small compared with the
changes observed in the internal flow structure.

\begin{figure}
    \centering
    \includegraphics[width=1\linewidth]{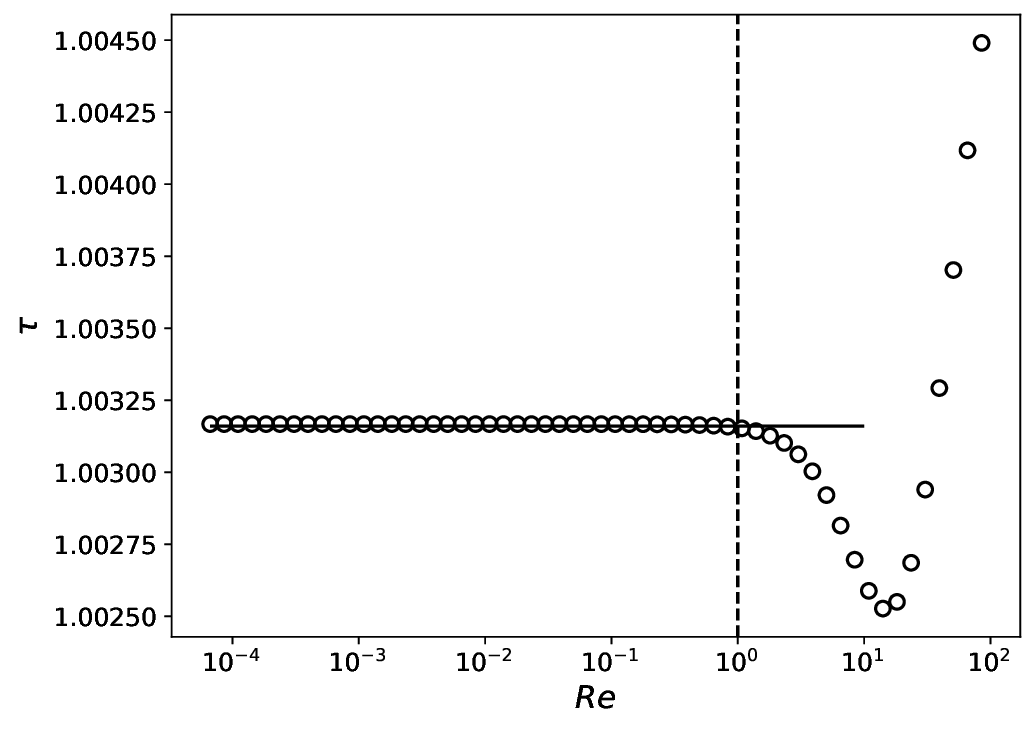}
    \caption{$\tau$ versus $Re$. The solid horizontal line marks the Stokes-limit ($Re\to0$) value $\tau_S=1.00316$ obtained from the semi-analytical solution of Eq.~\eqref{eq:stokes} (Sec.~\ref{sec:materials}).}
    \label{fig:tau}
\end{figure}

To quantify the changes in the flow structure observed in
Fig.~\ref{fig:velocity-field}, we calculated
the tortuosity $\tau$, which we defined as

\begin{equation}
\tau =
\frac{\langle |\mathbf{u}| \rangle_\Omega}
     {\langle u_x \rangle_\Omega},
\label{eq:tau}
\end{equation}

where $\langle |\mathbf{u}| \rangle$ is the area-weighted average
velocity magnitude and $\langle u_x \rangle$ is the area-weighted
streamwise velocity component \cite{Duda11, matyka2017granica}. For the present two-dimensional domain,
Eq.~\eqref{eq:tau} can equivalently be written as

\begin{equation}
\tau =
\frac{\sum_{i=1}^{N} |\mathbf{u}_i| A_i}
     {\sum_{i=1}^{N} u_{x,i} A_i},
\end{equation}

where $A_i$ is the area of the $i$th computational cell \cite{Naqvi26}. A value of
$\tau$ close to unity corresponds to a predominantly streamwise flow,
whereas larger values indicate stronger deviations from the main flow
direction and an increasing contribution of recirculating motion.

The variation of $\tau$ with Reynolds number is shown in
Fig.~\ref{fig:tau}. At very low $Re$, the tortuosity remains nearly
constant at $\tau\approx1.0032$, in agreement with the Stokes-limit value $\tau_S=1.00316$ obtained from the semi-analytical solution (Sec.~\ref{sec:materials}; solid horizontal line in Fig.~\ref{fig:tau}). The simulations follow this value up to $Re\approx1$, consistent with the onset of deviations from the Darcy relation in Fig.~\ref{fig:FF}. This is consistent with the velocity fields and streamlines
shown previously: viscous effects dominate, and the structure of the
flow is mainly imposed by the Koch geometry. Although the streamlines
follow the local variations of the boundary and small recirculation
pockets are already present, increasing the velocity within this range
does not significantly modify the overall flow paths.

With increasing $Re$, $\tau$ begins to decrease and reaches a minimum
of $\tau_{\min}\approx1.0025$ at $Re\approx14$. This decrease is consistent
with the streamline patterns in the lower panels of Fig.~\ref{fig:velocity-field}, where the
main through-flow becomes progressively straighter. As inertia
increases, the fluid tends to preserve its streamwise momentum and
becomes less sensitive to the smaller-scale variations of the Koch
boundary. The resulting straightening of the main flow paths reduces
the tortuosity. A similar decrease of $\tau$ with increasing inertia
has previously been observed in disordered and natural porous media
\cite{sniezek2024inertia,Naqvi26}.

Above the minimum, the trend reverses, and $\tau$ increases rapidly with
$Re$, reaching $\tau\approx1.0049$ at the largest $Re$ considered. This behavior is again consistent with the streamline results:
at higher Reynolds numbers, flow separation becomes stronger, and a
large recirculation region develops inside the Koch leaf cavity, together
with smaller secondary vortices. These recirculating structures
increase the contribution of transverse and backward velocities to
$|\mathbf{u}|$, while contributing less to the net streamwise
transport, and therefore increase $\tau$. The non-monotonic behavior
of tortuosity can thus be understood as the result of two competing
inertial mechanisms: straightening of the main through-flow, which
decreases $\tau$, and the growth of separated and vortical regions,
which increases it. The same trend was reported in Ref.~\cite{Naqvi26}, while inertia-induced changes
in tortuosity and energy distribution have also been observed in
disordered porous systems \cite{andrade1999inertial,sniezek2024inertia}.

The tortuosity result therefore provides a quantitative confirmation
of the qualitative changes already visible in the velocity and
streamline fields. At intermediate $Re$, inertia primarily straightens
the main flow, whereas at larger $Re$, the development of recirculation
becomes dominant and causes the rapid increase in $\tau$.
\begin{figure}
    \centering
    \includegraphics[width=1\linewidth]{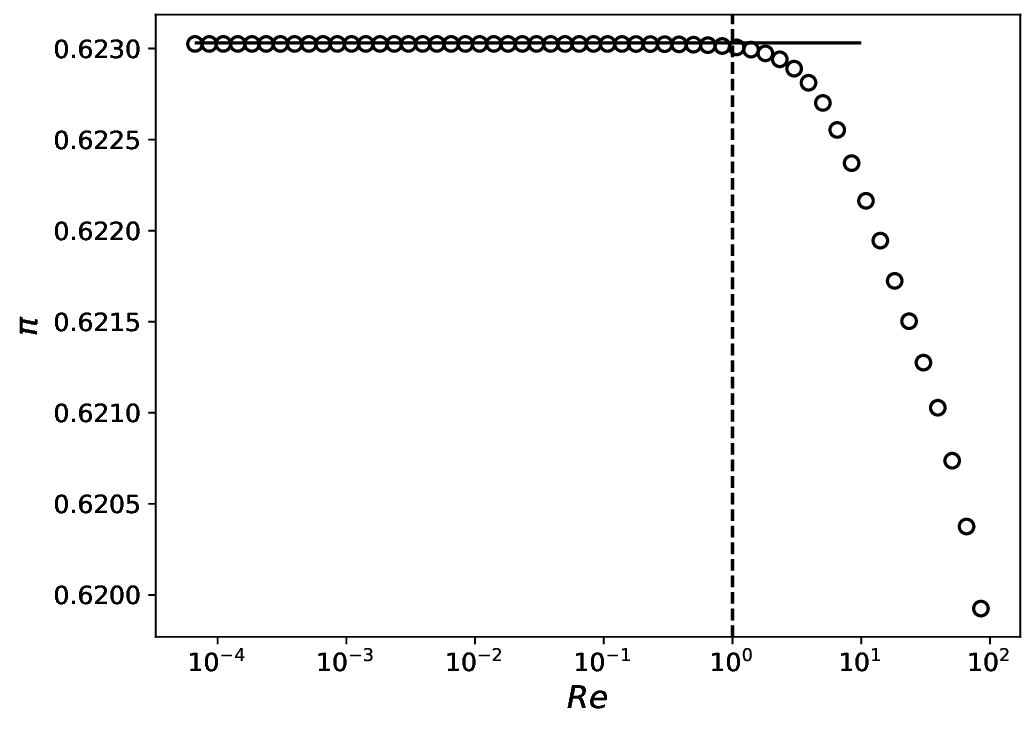}
    \caption{$\pi$ versus $Re$. The solid horizontal line marks the Stokes-limit value $\pi_S=0.62303$.}
    \label{fig:pi}
\end{figure}

To understand the mechanism behind the changes in tortuosity, we next
investigated the spatial distribution of kinetic energy. For this purpose,
we calculated the participation number $\pi$, which quantifies the degree
of localization of kinetic energy in the flow
\cite{andrade1999inertial,sniezek2024inertia,Naqvi26}. For the present
two-dimensional domain, the area-weighted participation number is defined
as

\begin{equation}
\pi =
\left(
A\sum_{i=1}^{N} q_i^2 A_i
\right)^{-1},
\label{eq:participation}
\end{equation}

where

\begin{equation}
q_i =
\frac{e_i}
{\displaystyle\sum_{j=1}^{N} e_j A_j},
\end{equation}

$A_i$ is the area of the $i$th computational cell,
$A=\sum_{i=1}^{N}A_i$ is the total fluid area, and the local kinetic-energy
density satisfies

\begin{equation}
e_i \propto |\mathbf{u}_i|^2.
\end{equation}

Since the density is constant, the factor $\rho/2$ in the kinetic-energy
density does not affect the value of $\pi$. A larger value of $\pi$
corresponds to a more spatially distributed kinetic-energy field, whereas
a decrease in $\pi$ indicates a stronger localization of kinetic energy.

The variation of $\pi$ with Reynolds number is shown in
Fig.~\ref{fig:pi}. At low $Re$, $\pi$ remains nearly constant at $\pi\approx0.6230$, coinciding with the Stokes-limit value $\pi_S=0.62303$ (solid line in Fig.~\ref{fig:pi}), indicating
that the relative spatial distribution of kinetic energy is essentially
unchanged in the viscous-dominated regime. As $Re$ increases, however,
$\pi$ decreases progressively, reaching $\pi\approx0.6199$ at the largest $Re$ considered. This behavior is opposite to that observed
in disordered and natural porous media, where increasing $Re$ was found
to increase $\pi$ as kinetic energy became distributed over a larger
fraction of the pore space
\cite{andrade1999inertial,sniezek2024inertia,Naqvi26}. For example,
Andrade \textit{et al.} reported a constant $\pi\approx0.37$ up to
$Re\approx0.3$, followed by a gradual increase of $\pi$ above this
transition \cite{andrade1999inertial}, and a similar rise of $\pi$ with
$Re$ was found by us for a disordered medium with porosity $\varphi=0.9$
\cite{sniezek2024inertia}. In the present Koch
geometry, increasing inertia therefore produces a stronger localization
of kinetic energy rather than a broader spatial distribution.

This behavior is consistent with the velocity and streamline fields shown
previously. With increasing $Re$, the main through-flow becomes straighter
and less sensitive to the small-scale variations of the Koch leaf boundary,
while a dominant recirculation region develops inside the cavity. The
flow is therefore increasingly organized into a more energetic main
stream and localized recirculating structures. Consequently, a smaller
fraction of the domain carries a larger fraction of the total kinetic
energy, leading to the observed decrease of $\pi$.

The decrease in $\pi$ also provides further insight into the
non-monotonic behavior of $\tau$. In the intermediate Reynolds-number
range, the straightening of the main flow reduces $\tau$, while the kinetic
energy simultaneously becomes more localized, as indicated by the
decrease in $\pi$. At larger $Re$, $\tau$ starts to increase because the
contribution of recirculating motion becomes sufficiently strong, whereas
$\pi$ continues to decrease. Thus, in the present geometry, tortuosity and
participation number do not follow the same trend. This differs from the
natural porous-medium case, where an increase in $\pi$ accompanied the
high-$Re$ increase in tortuosity \cite{Naqvi26}.
\begin{figure}
    \centering
    \includegraphics[width=1\linewidth]{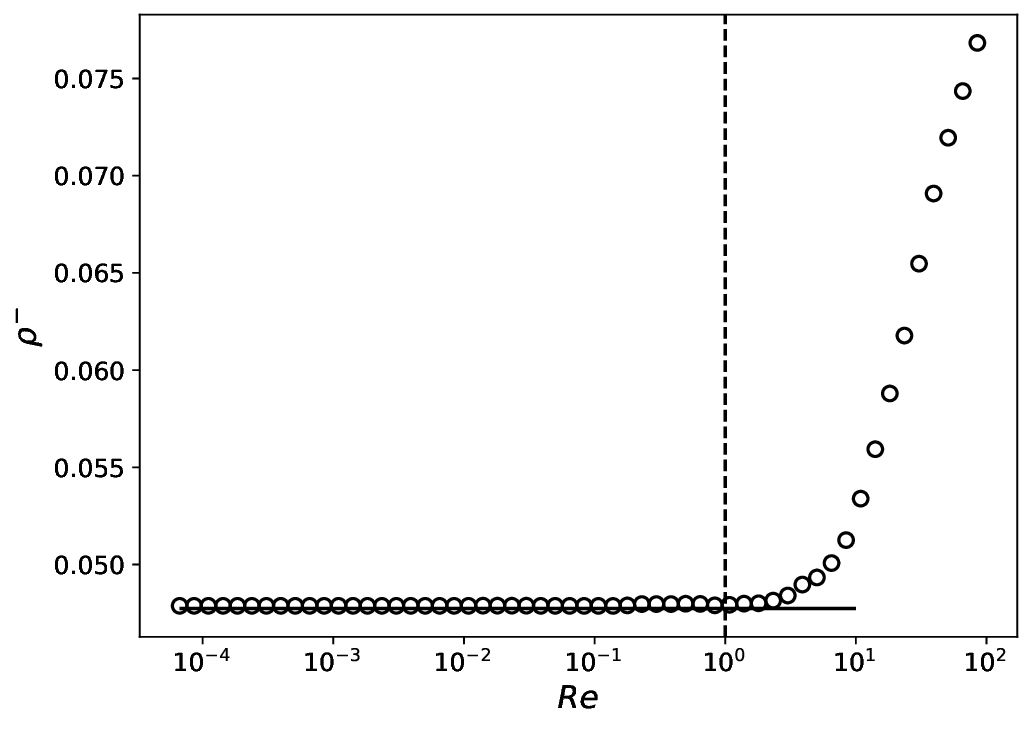}
    \caption{$\rho^{-}$ versus $Re$. The solid horizontal line marks the Stokes-limit value $\rho^{-}_S=0.0477$.}
    \label{fig:rho}
\end{figure}
To examine the contribution of recirculating motion more directly, we
calculated the backward-flow fraction $\rho^{-}$, which measures the
fraction of the fluid domain in which the streamwise velocity is negative
\cite{sniezek2024inertia,Naqvi26}. For the present two-dimensional domain,
the area-weighted backward-flow fraction is defined as

\begin{equation}
\rho^{-}
=
\frac{1}{A}
\sum_{i=1}^{N}
f(u_{x,i}) A_i,
\label{eq:rho-minus}
\end{equation}

where $A$ is the total fluid area, $A_i$ is the area of the $i$th cell,
and $u_{x,i}$ is the local streamwise velocity component. The indicator
function $f(u_x)$ is defined as

\begin{equation}
f(u_x)
=
\begin{cases}
1, & u_x < 0,\\
0, & \text{otherwise}.
\end{cases}
\label{eq:velocity-indicator}
\end{equation}

Thus, $\rho^{-}=0$ corresponds to a flow with no backward streamwise
motion, whereas an increasing $\rho^{-}$ indicates that a larger fraction
of the fluid domain is occupied by reverse flow.

The variation of $\rho^{-}$ with Reynolds number is shown in
Fig.~\ref{fig:rho}. At very low $Re$, $\rho^{-}$ remains nearly constant
at approximately $0.048$, in agreement with the Stokes-limit value $\rho^{-}_S=0.0477$ (solid line in Fig.~\ref{fig:rho}), that is $56\%$ of the Koch-leaf cavity area. This non-zero value is consistent with the
streamline field shown in Fig.~\ref{fig:velocity-field}, where small
recirculation pockets are already present around the finer geometrical
features of the Koch leaf boundary, even in the viscous-dominated regime \cite{moffatt1964viscous}.

As $Re$ increases, $\rho^{-}$ begins to rise, indicating a progressive
increase in the area occupied by backward flow. The increase becomes much
stronger at higher Reynolds numbers, and $\rho^{-}$ reaches approximately $0.077$, that is $89.5\%$ of the Koch-leaf cavity, at
the largest $Re$ considered. This trend is directly consistent with the
streamline results, where the initially small and localized recirculation
regions develop into a dominant large-scale vortex inside the Koch
cavity. The growth of $\rho^{-}$ therefore provides a quantitative
measure of the increasing importance of inertia-induced flow separation
and recirculation.

The relationship between $\rho^{-}$ and $\tau$ also helps to clarify the
two competing mechanisms identified above. In the intermediate
Reynolds-number range, $\rho^{-}$ already increases while $\tau$
decreases. This indicates that the growth of recirculation occurs
simultaneously with a stronger straightening of the main through-flow.
At this stage, the straightening effect dominates the global tortuosity.
At larger $Re$, the growth of the backward-flow region becomes
sufficiently rapid that $\tau$ reverses its trend and begins to increase.
The simultaneous increase of $\rho^{-}$ and $\tau$ at high $Re$ therefore
shows that recirculating motion becomes increasingly important in
determining the overall flow structure.

Together with the decrease of $\pi$, the growth of $\rho^{-}$ shows that the recirculating region inside the Koch leaf cavity expands while carrying only a small share of the kinetic energy, which becomes increasingly concentrated in the main through-flow.
\begin{figure}
    \centering
    \includegraphics[width=1\linewidth]{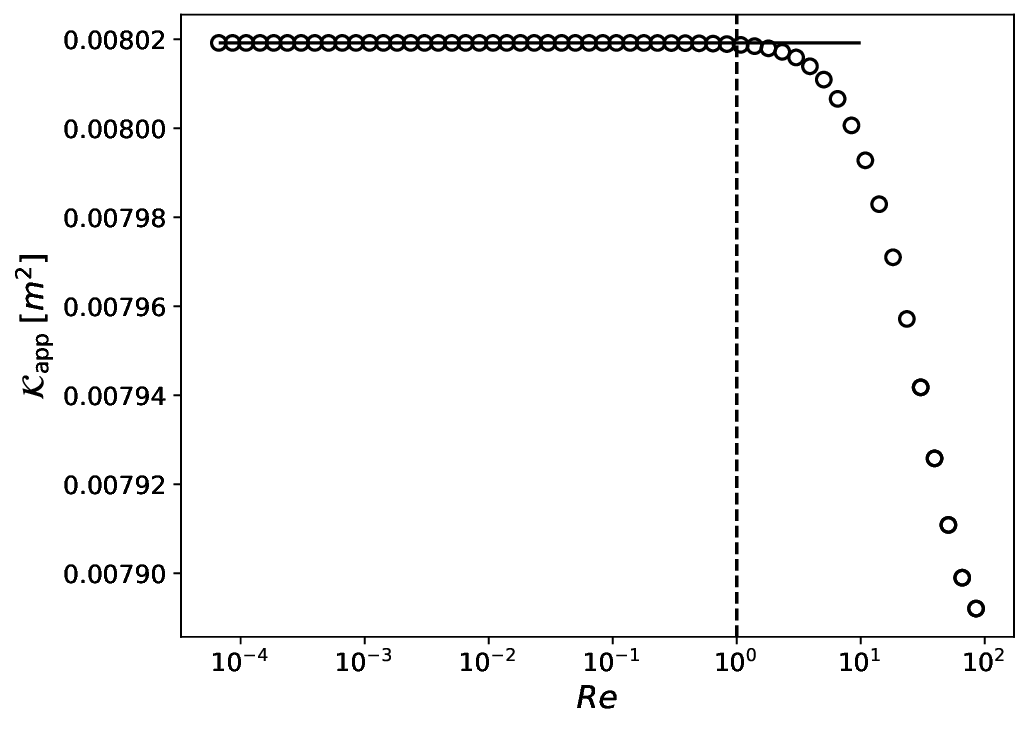}
    \caption{$\mathcal{K}_{app}$ versus $Re$. The solid horizontal line marks the Stokes-limit value $\mathcal{K}_S=8.019\times10^{-3}\,\mathrm{m}^2$.}
    \label{fig:kappa}
\end{figure}

Finally, we calculated the apparent permeability
$\mathcal{K}_{\mathrm{app}}$ from the Darcy relation as

\begin{equation}
\mathcal{K}_{\mathrm{app}}
=
\frac{\nu U L_p}{\Delta p},
\label{eq:kapp}
\end{equation}

where $\nu$ is the kinematic viscosity, $U$ is the mean inlet
velocity, and $L_p$ is the streamwise length over which the pressure drop
$\Delta p$ is measured.

The variation of $\mathcal{K}_{\mathrm{app}}$ with Reynolds number is
shown in Fig.~\ref{fig:kappa}. At low $Re$,
$\mathcal{K}_{\mathrm{app}}$ remains nearly constant at $\mathcal{K}_{\mathrm{app}}\approx8.02\times10^{-3}\,\mathrm{m}^2$, matching the Stokes-limit value $\mathcal{K}_{S}=8.019\times10^{-3}\,\mathrm{m}^2$ (solid line in Fig.~\ref{fig:kappa}), indicating that the pressure drop scales linearly with the flow velocity and that the system
is dominated by viscous effects. With increasing $Re$,
$\mathcal{K}_{\mathrm{app}}$ gradually decreases, reaching $\approx7.89\times10^{-3}\,\mathrm{m}^2$ at the largest $Re$ (a reduction of about $1.6\%$), showing the departure
from the Darcy regime and the increasing contribution of inertial
resistance.

The decrease becomes more pronounced at higher Reynolds numbers and is
consistent with the simultaneous growth of recirculation and tortuosity
observed previously. Thus, although the geometrical permeability is
unchanged, the apparent permeability decreases as inertia modifies the internal flow structure. A similar decrease of apparent permeability
with increasing $Re$ was reported for the low-porosity geological porous medium
\cite{Naqvi26}.

\begin{figure}
    \centering
    \includegraphics[width=1\linewidth]{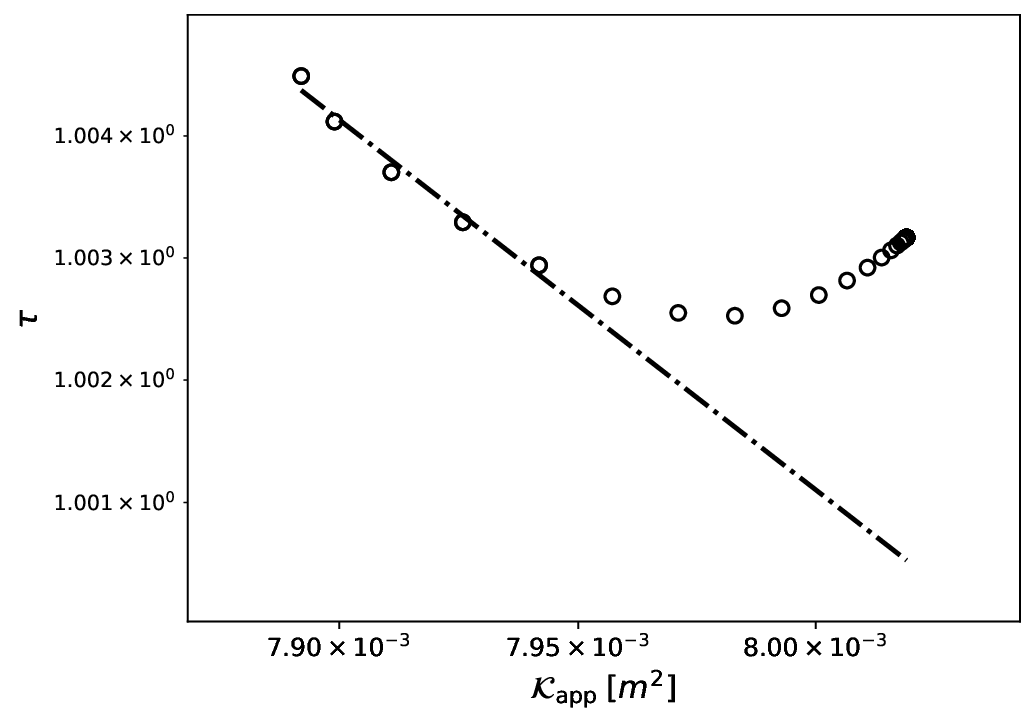}
    \caption{$\tau$ versus $\mathcal{K}_{app}$. The five highest-$Re$ points in the inertial regime are used for fitting, which represents the power law $\tau\propto\mathcal{K}_{app}^{-b}$, with $b\approx0.24$.}
    \label{fig:tk}
\end{figure}

To relate the changes in flow structure to the apparent permeability response, we examined the correlation between the tortuosity $\tau$
and the apparent permeability $\mathcal{K}_{app}$ (Fig.~\ref{fig:tk}).
Both quantities vary only weakly over the investigated range of
Reynolds numbers, consistent with the single, deterministic, multiscale
Koch leaf boundary probed here; nevertheless, their
joint variation follows the same qualitative trend reported for a
natural, low-porosity rock sample~\cite{Naqvi26}: as inertial
effects strengthen, the apparent permeability decreases while the
tortuosity increases, so that $\tau$ and $\mathcal{K}_{app}$ remain
inversely correlated throughout the inertial regime.

In the inertial range (i.e., beyond the minimum of
$\tau(Re)$ discussed above), this correlation can be described by a
power law,
\begin{equation}
    \tau \;\propto\; \mathcal{K}_{app}^{-b}.
    \label{eq:tau_kapp_koch}
\end{equation}
A fit with a free exponent to the five highest-$Re$ points in Fig.~\ref{fig:tk}
gives $b\approx0.24$ ($R^2 = 0.98$). Over this narrow range, the same points are equally
well reproduced by a linear form,
$\tau = 1.247 - 30.8\,\mathcal{K}_{app}$ (with $\mathcal{K}_{app}$
in $m^2$, $R^2 = 0.98$); the power-law and linear descriptions are locally
indistinguishable because $\tau$ and $\mathcal{K}_{app}$ vary by only
$0.15\%$ and $0.6\%$, respectively, across this inertial-regime data range.
Because the fit is based on only five points spanning such a narrow range,
the power-law form and the value of $b$ should be regarded as indicative
rather than conclusive; establishing the functional form of the
$\tau$--$\mathcal{K}_{app}$ relation unambiguously would require
additional simulations at higher $Re$. The exponent obtained for the
Koch-leaf geometry is nevertheless comparable to that of the natural rock
sample \cite{Naqvi26}, even though the relative variation of $\tau$ and
$\mathcal{K}_{app}$ in the present system is roughly two orders of magnitude
smaller (below $2\%$ here versus tens of percent in Ref.~\cite{Naqvi26}).
We attribute this small variation to the single, isolated Koch leaf boundary
probed here: inertia-driven flow redistribution only mildly affects the
streamwise resistance, unlike in a low-porosity medium. That a comparable
exponent appears despite this difference in scale suggests that the local
form of the $\tau$--$\mathcal{K}_{app}$ coupling is set by the inertial flow
itself rather than by the degree of pore-space confinement.

Taken together with the non-monotonic behavior of $\tau(Re)$
(Fig.~\ref{fig:tau}) and the monotonic decrease of
$\mathcal{K}_{app}(Re)$ (Fig.~\ref{fig:kappa}), this result
indicates that the coupling between tortuosity and apparent
permeability identified for disordered, natural porous media persists
even in a fully deterministic, highly open geometry such as the Koch
leaf. It is therefore not a consequence of pore-scale disorder but
of the more general inertia-induced change in the flow structure---the localization
of kinetic energy and the growth of recirculating, backward-flow regions---that accompanies the onset of inertial effects.

\begin{figure}
    \centering
     \includegraphics[width=1\linewidth]{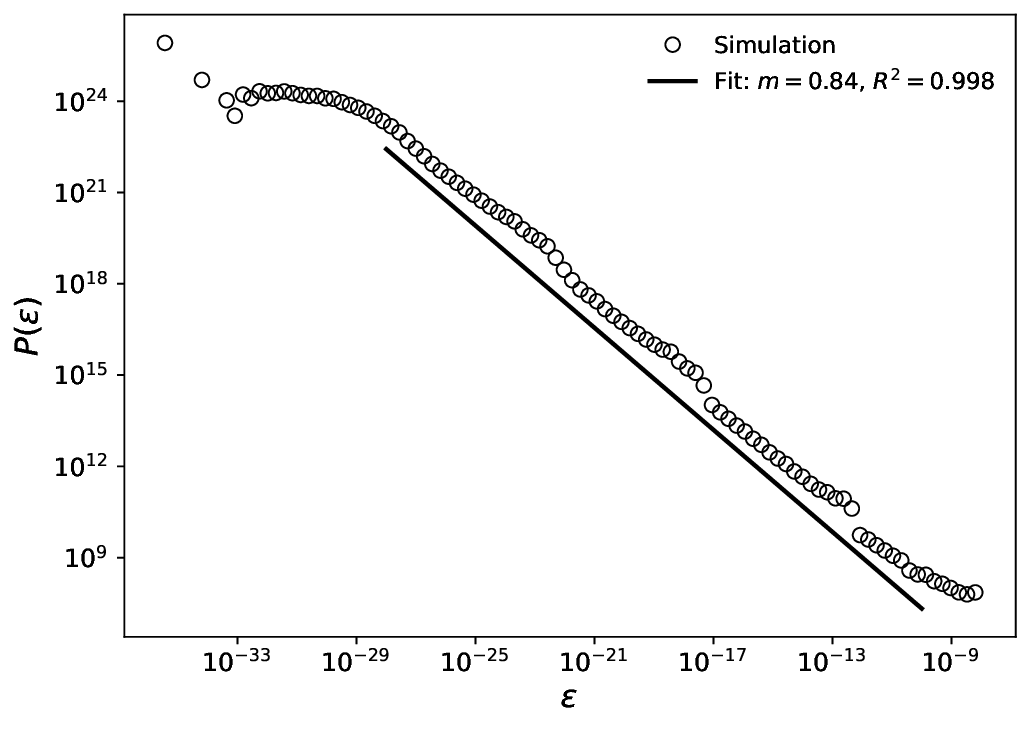}\\
     \includegraphics[width=1\linewidth]{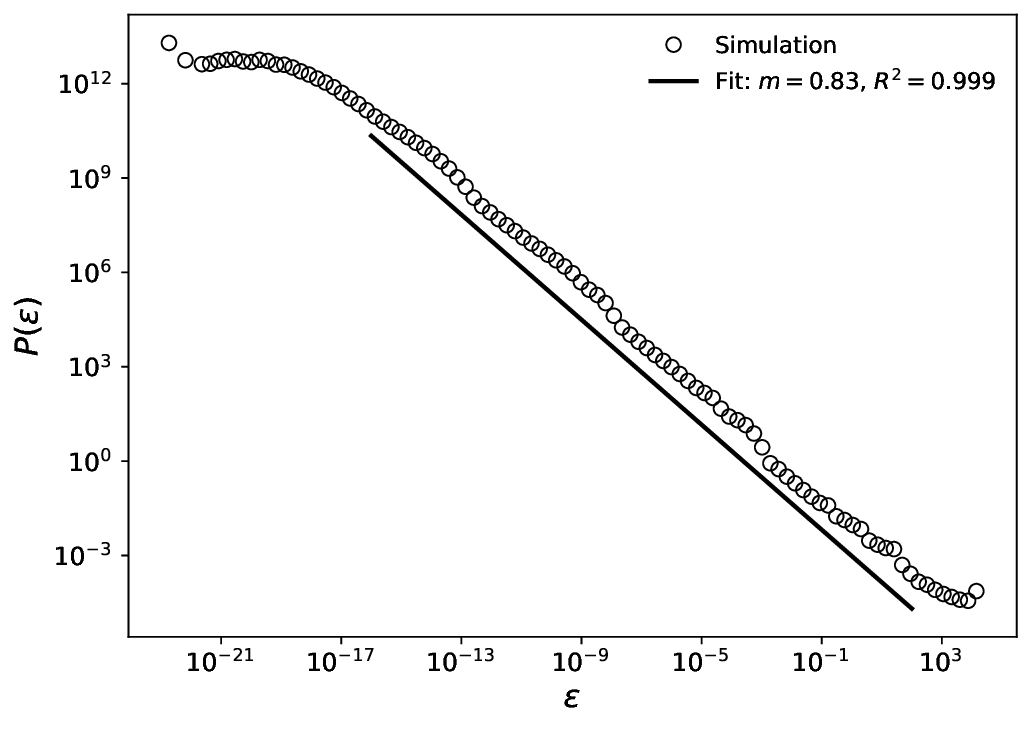}
    \caption{PDF of the local kinetic energy for low- ($Re=6.67\times10^{-5}$ (top)) and high- ($Re\approx85.09$ (bottom)) Reynolds numbers. The solid line $\propto\varepsilon^{-m}$, is shifted downward by a factor of 10 as a guide to the eye.}
    \label{fig:pdf}
\end{figure}

To complement the participation number, which measures how kinetic energy is spread in space, we examined the probability distribution function of the local kinetic energy, $\varepsilon = u^2+v^2$, at a representative low, $Re=6.67\times10^{-5}$, and high, $Re\approx85.09$, Reynolds number (Fig.~\ref{fig:pdf}). At both Reynolds numbers, $P(\varepsilon)$ follows a power law over several decades,
\begin{equation}
    P(\varepsilon) \;\propto\; \varepsilon^{-m},
    \label{eq:energy_pdf}
\end{equation}
with $m=0.84$ at low $Re$ and $m=0.83$ at high $Re$. Because $\varepsilon\propto U^2$, the energies at the two Reynolds numbers differ by about twelve orders of magnitude: the mean energy $\langle\varepsilon\rangle$ increases by a factor of $1.6\times10^{12}$, close to $(Re_{\mathrm{high}}/Re_{\mathrm{low}})^2$. The fits were therefore performed over the same range of normalized energy, $\varepsilon/\langle\varepsilon\rangle\approx10^{-16}$--$10^{-10}$, that is, over the same part of each distribution. Within this range, the exponent is unaffected by inertia. When rescaled by $\langle\varepsilon\rangle$, the two distributions collapse onto each other over about twenty decades and differ only at their ends: at the lowest energies, $\varepsilon/\langle\varepsilon\rangle\lesssim10^{-21}$, the high-$Re$ distribution lies higher, indicating a larger fraction of nearly stagnant fluid, while at the highest energies, $\varepsilon/\langle\varepsilon\rangle\approx10^{-6}$--$1$, it is slightly shallower (local slope $\approx0.70$ compared with $\approx0.76$ at low $Re$).

A power-law energy distribution of this type was first reported by Andrade~\textit{et al.}~\cite{andrade1997fluid} for two-dimensional flow through a disordered percolation cluster, where $n(E)\propto E^{-\alpha}$ with $\alpha = 0.90 \pm 0.03$ for the full cluster and $\alpha = 0.64 \pm 0.05$ for the flow-carrying backbone alone. The exponent obtained for the Koch leaf, $m\approx0.85$, is close to the full-cluster value, which includes the stagnant regions, suggesting that in both systems the power law is set by slowly moving fluid near the solid boundary. In contrast to the percolation cluster, where the exponent decreased to $\alpha = 0.79 \pm 0.03$ at high $Re$ as vortices and zones of flow separation appeared \cite{andrade1997fluid}, $m$ does not change in the Koch-leaf geometry over the investigated range. We attribute this to the origin of the power-law range: it is formed by the slow fluid in the small-scale recesses of the Koch leaf boundary, where the local flow remains viscous and of Moffatt type \cite{moffatt1964viscous} even at the highest $Re$ considered, whereas inertia acts mainly on the main through-flow and on the large primary vortex in the cavity.

This also clarifies the relation between the energy distribution and the participation number. In Ref.~\cite{andrade1999inertial} and in the natural porous medium of Ref.~\cite{Naqvi26}, $\pi$ increases with $Re$, reflecting the spatial delocalization of kinetic energy as new flow channels open up. In the Koch-leaf geometry, by contrast, $\pi$ decreases with $Re$ (Fig.~\ref{fig:pi}) while $m$ remains unchanged. The decrease in $\pi$ therefore does not originate from the power-law part of $P(\varepsilon)$, but from its ends. Because $\pi$ depends on $e_i^2$ [Eq.~\eqref{eq:participation}], it is dominated by the high-energy cells, and the change in the high-energy tail, together with the growth of nearly stagnant fluid at the low-energy end, reflects the increasing contrast between the faster, straighter main through-flow and the expanding recirculation region inside the cavity. Inertia thus concentrates kinetic energy into the main stream without altering the scale-free statistics of the slow, near-wall flow set by the Koch leaf geometry.

\section{CONCLUSIONS}
We studied inertial effects in steady, laminar flow through a channel with a sixth-order Koch-leaf cavity ($\varphi\approx1$) by means of OpenFOAM simulations covering $6.67\times10^{-5}\le Re\le85.09$. The simulations were validated against a semi-analytical solution for Stokes regime built from Moffatt corner modes, which reproduces the low-$Re$ values of $\tau$, $\pi$, $\rho^{-}$, and $\mathcal{K}_{\mathrm{app}}$. The macroscopic response hides the transition: the friction factor deviates from Darcy's law by only $1.4\%$, and the apparent permeability decreases by $1.6\%$. The internal flow, by contrast, changes qualitatively. Recirculation pockets merge into a primary vortex, the backward-flow region grows from $56\%$ to $89.5\%$ of the cavity, and the tortuosity varies non-monotonically, with a minimum at $Re\approx14$, as flow straightening first dominates and recirculation later prevails. In contrast to the disordered porous media, we found that the participation number decreases with $Re$, as inertia concentrates kinetic energy in the main stream, whereas the local kinetic energy follows a power-law distribution with the same exponent, $m\approx0.83$, at the lowest and highest $Re$, with inertia altering only its tails. In the inertial regime, tortuosity and apparent permeability are related by a power law with an exponent comparable to that of natural rock \cite{Naqvi26}. These results show that key inertial effects known from disordered porous media also arise in a single deterministic geometry, making fractal geometries a controlled testbed for separating the effects of inertia from those of the structure.

\section*{Acknowledgment}
Funded by National Science Centre, Poland under the OPUS call in the Weave programme 2021/43/I/ST3/00228.
This research was funded in whole or in part by National Science Centre (2021/43/I/ST3/00228). For the purpose of Open Access,
the author has applied a CC-BY public copyright licence to any Author Accepted Manuscript (AAM) version arising from this submission. M. Matyka acknowledges the financial support from the Slovenian Research And Innovation Agency (ARIS) research core funding No. P2-0095.


\begin{thebibliography}{38}%
\makeatletter
\providecommand \@ifxundefined [1]{%
 \@ifx{#1\undefined}
}%
\providecommand \@ifnum [1]{%
 \ifnum #1\expandafter \@firstoftwo
 \else \expandafter \@secondoftwo
 \fi
}%
\providecommand \@ifx [1]{%
 \ifx #1\expandafter \@firstoftwo
 \else \expandafter \@secondoftwo
 \fi
}%
\providecommand \natexlab [1]{#1}%
\providecommand \enquote  [1]{``#1''}%
\providecommand \bibnamefont  [1]{#1}%
\providecommand \bibfnamefont [1]{#1}%
\providecommand \citenamefont [1]{#1}%
\providecommand \href@noop [0]{\@secondoftwo}%
\providecommand \href [0]{\begingroup \@sanitize@url \@href}%
\providecommand \@href[1]{\@@startlink{#1}\@@href}%
\providecommand \@@href[1]{\endgroup#1\@@endlink}%
\providecommand \@sanitize@url [0]{\catcode `\\12\catcode `\$12\catcode `\&12\catcode `\#12\catcode `\^12\catcode `\_12\catcode `\%12\relax}%
\providecommand \@@startlink[1]{}%
\providecommand \@@endlink[0]{}%
\providecommand \url  [0]{\begingroup\@sanitize@url \@url }%
\providecommand \@url [1]{\endgroup\@href {#1}{\urlprefix }}%
\providecommand \urlprefix  [0]{URL }%
\providecommand \Eprint [0]{\href }%
\providecommand \doibase [0]{https://doi.org/}%
\providecommand \selectlanguage [0]{\@gobble}%
\providecommand \bibinfo  [0]{\@secondoftwo}%
\providecommand \bibfield  [0]{\@secondoftwo}%
\providecommand \translation [1]{[#1]}%
\providecommand \BibitemOpen [0]{}%
\providecommand \bibitemStop [0]{}%
\providecommand \bibitemNoStop [0]{.\EOS\space}%
\providecommand \EOS [0]{\spacefactor3000\relax}%
\providecommand \BibitemShut  [1]{\csname bibitem#1\endcsname}%
\let\auto@bib@innerbib\@empty
\bibitem [{\citenamefont {Hyman}(2020)}]{hyman2020flow}%
  \BibitemOpen
  \bibfield  {author} {\bibinfo {author} {\bibfnamefont {J.~D.}\ \bibnamefont {Hyman}},\ }\bibfield  {title} {\bibinfo {title} {Flow channeling in fracture networks: Characterizing the effect of density on preferential flow path formation},\ }\href@noop {} {\bibfield  {journal} {\bibinfo  {journal} {Water Resources Research}\ }\textbf {\bibinfo {volume} {56}},\ \bibinfo {pages} {e2020WR027986} (\bibinfo {year} {2020})}\BibitemShut {NoStop}%
\bibitem [{\citenamefont {Andrade}\ \emph {et~al.}(1999)\citenamefont {Andrade}, \citenamefont {Costa}, \citenamefont {Almeida}, \citenamefont {Makse},\ and\ \citenamefont {Stanley}}]{andrade1999inertial}%
  \BibitemOpen
  \bibfield  {author} {\bibinfo {author} {\bibfnamefont {J.~S.~J.}\ \bibnamefont {Andrade}}, \bibinfo {author} {\bibfnamefont {U.~M.~S.}\ \bibnamefont {Costa}}, \bibinfo {author} {\bibfnamefont {M.~P.}\ \bibnamefont {Almeida}}, \bibinfo {author} {\bibfnamefont {H.~A.}\ \bibnamefont {Makse}},\ and\ \bibinfo {author} {\bibfnamefont {H.~E.}\ \bibnamefont {Stanley}},\ }\bibfield  {title} {\bibinfo {title} {Inertial effects on fluid flow through disordered porous media},\ }\href {https://doi.org/10.1103/PhysRevLett.82.5249} {\bibfield  {journal} {\bibinfo  {journal} {Physical Review Letters}\ }\textbf {\bibinfo {volume} {82}},\ \bibinfo {pages} {5249} (\bibinfo {year} {1999})}\BibitemShut {NoStop}%
\bibitem [{\citenamefont {Hovorka}\ \emph {et~al.}(2019)\citenamefont {Hovorka}, \citenamefont {Trevi{\~n}o}, \citenamefont {Nicot} \emph {et~al.}}]{hovorka2019dynamic}%
  \BibitemOpen
  \bibfield  {author} {\bibinfo {author} {\bibfnamefont {S.~C.}\ \bibnamefont {Hovorka}}, \bibinfo {author} {\bibfnamefont {R.~H.}\ \bibnamefont {Trevi{\~n}o}}, \bibinfo {author} {\bibfnamefont {M.~J.}\ \bibnamefont {Nicot}}, \emph {et~al.},\ }\bibfield  {title} {\bibinfo {title} {Dynamic processes of {CO$_2$} storage in the field: 1. multiscale and multiphase flow behavior},\ }\href {https://doi.org/10.1029/2019WR025688} {\bibfield  {journal} {\bibinfo  {journal} {Water Resources Research}\ }\textbf {\bibinfo {volume} {55}},\ \bibinfo {pages} {1} (\bibinfo {year} {2019})}\BibitemShut {NoStop}%
\bibitem [{\citenamefont {Matyka}\ and\ \citenamefont {M{\k{a}}drala}(2017)}]{matyka2017granica}%
  \BibitemOpen
  \bibfield  {author} {\bibinfo {author} {\bibfnamefont {M.}~\bibnamefont {Matyka}}\ and\ \bibinfo {author} {\bibfnamefont {M.}~\bibnamefont {M{\k{a}}drala}},\ }\bibfield  {title} {\bibinfo {title} {Granica laminarnego przep{\l}ywu wody w obsypce {\.z}wirowej studni eksploatacyjnej},\ }\href@noop {} {\bibfield  {journal} {\bibinfo  {journal} {Przegl{\k{a}}d Geologiczny}\ }\textbf {\bibinfo {volume} {65}},\ \bibinfo {pages} {1334} (\bibinfo {year} {2017})}\BibitemShut {NoStop}%
\bibitem [{\citenamefont {Cao}\ \emph {et~al.}(2019)\citenamefont {Cao}, \citenamefont {Jung},\ and\ \citenamefont {Radonjic}}]{cao2019application}%
  \BibitemOpen
  \bibfield  {author} {\bibinfo {author} {\bibfnamefont {S.~C.}\ \bibnamefont {Cao}}, \bibinfo {author} {\bibfnamefont {J.}~\bibnamefont {Jung}},\ and\ \bibinfo {author} {\bibfnamefont {M.}~\bibnamefont {Radonjic}},\ }\bibfield  {title} {\bibinfo {title} {Application of microfluidic pore models for flow, transport, and reaction in geological porous media: From a single test bed to multifunction real-time analysis tool},\ }\href@noop {} {\bibfield  {journal} {\bibinfo  {journal} {Microsystem Technologies}\ }\textbf {\bibinfo {volume} {25}},\ \bibinfo {pages} {4035} (\bibinfo {year} {2019})}\BibitemShut {NoStop}%
\bibitem [{\citenamefont {Banerjee}\ and\ \citenamefont {Paul}(2021)}]{banerjee2021developments}%
  \BibitemOpen
  \bibfield  {author} {\bibinfo {author} {\bibfnamefont {A.}~\bibnamefont {Banerjee}}\ and\ \bibinfo {author} {\bibfnamefont {D.}~\bibnamefont {Paul}},\ }\bibfield  {title} {\bibinfo {title} {Developments and applications of porous medium combustion: A recent review},\ }\href@noop {} {\bibfield  {journal} {\bibinfo  {journal} {Energy}\ }\textbf {\bibinfo {volume} {221}},\ \bibinfo {pages} {119868} (\bibinfo {year} {2021})}\BibitemShut {NoStop}%
\bibitem [{\citenamefont {Zhang}\ \emph {et~al.}(2024)\citenamefont {Zhang}, \citenamefont {Zhou}, \citenamefont {Zhang}, \citenamefont {Zhang}, \citenamefont {Liu}, \citenamefont {Zhu}, \citenamefont {Sun}, \citenamefont {Li},\ and\ \citenamefont {Gu}}]{zhang2024optimal}%
  \BibitemOpen
  \bibfield  {author} {\bibinfo {author} {\bibfnamefont {Y.}~\bibnamefont {Zhang}}, \bibinfo {author} {\bibfnamefont {Y.}~\bibnamefont {Zhou}}, \bibinfo {author} {\bibfnamefont {Y.}~\bibnamefont {Zhang}}, \bibinfo {author} {\bibfnamefont {Z.}~\bibnamefont {Zhang}}, \bibinfo {author} {\bibfnamefont {J.}~\bibnamefont {Liu}}, \bibinfo {author} {\bibfnamefont {J.}~\bibnamefont {Zhu}}, \bibinfo {author} {\bibfnamefont {Z.}~\bibnamefont {Sun}}, \bibinfo {author} {\bibfnamefont {L.}~\bibnamefont {Li}},\ and\ \bibinfo {author} {\bibfnamefont {H.}~\bibnamefont {Gu}},\ }\bibfield  {title} {\bibinfo {title} {Optimal design of silver-loaded zeolite filter structure based on porous media model},\ }\href@noop {} {\bibfield  {journal} {\bibinfo  {journal} {Separation and Purification Technology}\ }\textbf {\bibinfo {volume} {334}},\ \bibinfo {pages} {125905} (\bibinfo {year} {2024})}\BibitemShut {NoStop}%
\bibitem [{\citenamefont {Wang}\ \emph {et~al.}(2014)\citenamefont {Wang}, \citenamefont {Shuai}, \citenamefont {Wang}, \citenamefont {Leng},\ and\ \citenamefont {Tan}}]{wang2014thermal}%
  \BibitemOpen
  \bibfield  {author} {\bibinfo {author} {\bibfnamefont {F.}~\bibnamefont {Wang}}, \bibinfo {author} {\bibfnamefont {Y.}~\bibnamefont {Shuai}}, \bibinfo {author} {\bibfnamefont {Z.}~\bibnamefont {Wang}}, \bibinfo {author} {\bibfnamefont {Y.}~\bibnamefont {Leng}},\ and\ \bibinfo {author} {\bibfnamefont {H.}~\bibnamefont {Tan}},\ }\bibfield  {title} {\bibinfo {title} {Thermal and chemical reaction performance analyses of steam methane reforming in porous media solar thermochemical reactor},\ }\href@noop {} {\bibfield  {journal} {\bibinfo  {journal} {International Journal of Hydrogen Energy}\ }\textbf {\bibinfo {volume} {39}},\ \bibinfo {pages} {718} (\bibinfo {year} {2014})}\BibitemShut {NoStop}%
\bibitem [{\citenamefont {Neuman}(1977)}]{neuman1977theoretical}%
  \BibitemOpen
  \bibfield  {author} {\bibinfo {author} {\bibfnamefont {S.~P.}\ \bibnamefont {Neuman}},\ }\bibfield  {title} {\bibinfo {title} {Theoretical derivation of darcy's law},\ }\href@noop {} {\bibfield  {journal} {\bibinfo  {journal} {Acta Mechanica}\ }\textbf {\bibinfo {volume} {25}},\ \bibinfo {pages} {153} (\bibinfo {year} {1977})}\BibitemShut {NoStop}%
\bibitem [{\citenamefont {Whitaker}(1986)}]{whitaker1986flow}%
  \BibitemOpen
  \bibfield  {author} {\bibinfo {author} {\bibfnamefont {S.}~\bibnamefont {Whitaker}},\ }\bibfield  {title} {\bibinfo {title} {Flow in porous media i: A theoretical derivation of darcy's law},\ }\href {https://doi.org/10.1007/BF01036523} {\bibfield  {journal} {\bibinfo  {journal} {Transport in Porous Media}\ }\textbf {\bibinfo {volume} {1}},\ \bibinfo {pages} {3} (\bibinfo {year} {1986})}\BibitemShut {NoStop}%
\bibitem [{\citenamefont {Whitaker}(1996)}]{whitaker1996Forchheimer}%
  \BibitemOpen
  \bibfield  {author} {\bibinfo {author} {\bibfnamefont {S.}~\bibnamefont {Whitaker}},\ }\bibfield  {title} {\bibinfo {title} {The forchheimer equation: A theoretical development},\ }\href {https://doi.org/10.1007/BF00141261} {\bibfield  {journal} {\bibinfo  {journal} {Transport in Porous Media}\ }\textbf {\bibinfo {volume} {25}},\ \bibinfo {pages} {27} (\bibinfo {year} {1996})}\BibitemShut {NoStop}%
\bibitem [{\citenamefont {Barree}\ and\ \citenamefont {Conway}(2004)}]{barree2004beyond}%
  \BibitemOpen
  \bibfield  {author} {\bibinfo {author} {\bibfnamefont {R.~D.}\ \bibnamefont {Barree}}\ and\ \bibinfo {author} {\bibfnamefont {M.~W.}\ \bibnamefont {Conway}},\ }\bibfield  {title} {\bibinfo {title} {Beyond beta factors: A complete model for darcy, forchheimer, and trans-forchheimer flow in porous media},\ }in\ \href@noop {} {\emph {\bibinfo {booktitle} {Proceedings of the 2004 SPE Annual Technical Conference and Exhibition}}}\ (\bibinfo  {publisher} {Society of Petroleum Engineers},\ \bibinfo {year} {2004})\BibitemShut {NoStop}%
\bibitem [{\citenamefont {Fourar}\ \emph {et~al.}(2005)\citenamefont {Fourar}, \citenamefont {Lenormand}, \citenamefont {Karimi-Fard},\ and\ \citenamefont {Horne}}]{fourar2005inertia}%
  \BibitemOpen
  \bibfield  {author} {\bibinfo {author} {\bibfnamefont {M.}~\bibnamefont {Fourar}}, \bibinfo {author} {\bibfnamefont {R.}~\bibnamefont {Lenormand}}, \bibinfo {author} {\bibfnamefont {M.}~\bibnamefont {Karimi-Fard}},\ and\ \bibinfo {author} {\bibfnamefont {R.~N.}\ \bibnamefont {Horne}},\ }\bibfield  {title} {\bibinfo {title} {Inertia effects in high-rate flow through heterogeneous porous media},\ }\href {https://doi.org/10.1007/s11242-004-6800-6} {\bibfield  {journal} {\bibinfo  {journal} {Transport in Porous Media}\ }\textbf {\bibinfo {volume} {60}},\ \bibinfo {pages} {353} (\bibinfo {year} {2005})}\BibitemShut {NoStop}%
\bibitem [{\citenamefont {Arthur}(2020)}]{arthur2020piv}%
  \BibitemOpen
  \bibfield  {author} {\bibinfo {author} {\bibfnamefont {J.}~\bibnamefont {Arthur}},\ }\bibfield  {title} {\bibinfo {title} {{PIV} study of flow through and over porous media at the onset of inertia},\ }\href {https://doi.org/10.1016/j.advwatres.2020.103793} {\bibfield  {journal} {\bibinfo  {journal} {Advances in Water Resources}\ }\textbf {\bibinfo {volume} {146}},\ \bibinfo {pages} {103793} (\bibinfo {year} {2020})}\BibitemShut {NoStop}%
\bibitem [{\citenamefont {Balhoff}\ and\ \citenamefont {Wheeler}(2009)}]{balhoff2009predictive}%
  \BibitemOpen
  \bibfield  {author} {\bibinfo {author} {\bibfnamefont {M.~T.}\ \bibnamefont {Balhoff}}\ and\ \bibinfo {author} {\bibfnamefont {M.~F.}\ \bibnamefont {Wheeler}},\ }\bibfield  {title} {\bibinfo {title} {A predictive pore-scale model for non-darcy flow in porous media},\ }\href {https://doi.org/10.2118/110838-PA} {\bibfield  {journal} {\bibinfo  {journal} {SPE Journal}\ }\textbf {\bibinfo {volume} {14}},\ \bibinfo {pages} {579} (\bibinfo {year} {2009})}\BibitemShut {NoStop}%
\bibitem [{\citenamefont {Zeng}\ and\ \citenamefont {Grigg}(2006)}]{zeng2006criterion}%
  \BibitemOpen
  \bibfield  {author} {\bibinfo {author} {\bibfnamefont {Z.}~\bibnamefont {Zeng}}\ and\ \bibinfo {author} {\bibfnamefont {R.}~\bibnamefont {Grigg}},\ }\bibfield  {title} {\bibinfo {title} {A criterion for non-darcy flow in porous media},\ }\href {https://doi.org/10.1007/s11242-005-2720-3} {\bibfield  {journal} {\bibinfo  {journal} {Transport in Porous Media}\ }\textbf {\bibinfo {volume} {63}},\ \bibinfo {pages} {57} (\bibinfo {year} {2006})}\BibitemShut {NoStop}%
\bibitem [{\citenamefont {Arbabi}\ and\ \citenamefont {Sahimi}(2024)}]{arbabi2024transition}%
  \BibitemOpen
  \bibfield  {author} {\bibinfo {author} {\bibfnamefont {S.}~\bibnamefont {Arbabi}}\ and\ \bibinfo {author} {\bibfnamefont {M.}~\bibnamefont {Sahimi}},\ }\bibfield  {title} {\bibinfo {title} {The transition from darcy to nonlinear flow in heterogeneous porous media: I---single-phase flow},\ }\href@noop {} {\bibfield  {journal} {\bibinfo  {journal} {Transport in Porous Media}\ }\textbf {\bibinfo {volume} {151}},\ \bibinfo {pages} {795} (\bibinfo {year} {2024})}\BibitemShut {NoStop}%
\bibitem [{\citenamefont {{\'S}nie{\.z}ek}\ \emph {et~al.}(2024)\citenamefont {{\'S}nie{\.z}ek}, \citenamefont {Naqvi},\ and\ \citenamefont {Matyka}}]{sniezek2024inertia}%
  \BibitemOpen
  \bibfield  {author} {\bibinfo {author} {\bibfnamefont {D.}~\bibnamefont {{\'S}nie{\.z}ek}}, \bibinfo {author} {\bibfnamefont {S.~B.}\ \bibnamefont {Naqvi}},\ and\ \bibinfo {author} {\bibfnamefont {M.}~\bibnamefont {Matyka}},\ }\bibfield  {title} {\bibinfo {title} {Inertia onset in disordered porous media flow},\ }\href {https://doi.org/10.1103/PhysRevE.110.045103} {\bibfield  {journal} {\bibinfo  {journal} {Physical Review E}\ }\textbf {\bibinfo {volume} {110}},\ \bibinfo {pages} {045103} (\bibinfo {year} {2024})}\BibitemShut {NoStop}%
\bibitem [{\citenamefont {Muljadi}\ \emph {et~al.}(2016)\citenamefont {Muljadi}, \citenamefont {Blunt}, \citenamefont {Raeini},\ and\ \citenamefont {Bijeljic}}]{muljadi2016impact}%
  \BibitemOpen
  \bibfield  {author} {\bibinfo {author} {\bibfnamefont {B.~P.}\ \bibnamefont {Muljadi}}, \bibinfo {author} {\bibfnamefont {M.~J.}\ \bibnamefont {Blunt}}, \bibinfo {author} {\bibfnamefont {A.~Q.}\ \bibnamefont {Raeini}},\ and\ \bibinfo {author} {\bibfnamefont {B.}~\bibnamefont {Bijeljic}},\ }\bibfield  {title} {\bibinfo {title} {The impact of porous media heterogeneity on non-darcy flow behaviour from pore-scale simulation},\ }\href@noop {} {\bibfield  {journal} {\bibinfo  {journal} {Advances in Water Resources}\ }\textbf {\bibinfo {volume} {95}},\ \bibinfo {pages} {329} (\bibinfo {year} {2016})}\BibitemShut {NoStop}%
\bibitem [{\citenamefont {Nissan}\ and\ \citenamefont {Berkowitz}(2018)}]{nissan2018inertial}%
  \BibitemOpen
  \bibfield  {author} {\bibinfo {author} {\bibfnamefont {A.}~\bibnamefont {Nissan}}\ and\ \bibinfo {author} {\bibfnamefont {B.}~\bibnamefont {Berkowitz}},\ }\bibfield  {title} {\bibinfo {title} {Inertial effects on flow and transport in heterogeneous porous media},\ }\href {https://doi.org/10.1103/PhysRevLett.120.054504} {\bibfield  {journal} {\bibinfo  {journal} {Physical Review Letters}\ }\textbf {\bibinfo {volume} {120}},\ \bibinfo {pages} {054504} (\bibinfo {year} {2018})}\BibitemShut {NoStop}%
\bibitem [{\citenamefont {Basak}(1977)}]{basak1977non}%
  \BibitemOpen
  \bibfield  {author} {\bibinfo {author} {\bibfnamefont {P.}~\bibnamefont {Basak}},\ }\bibfield  {title} {\bibinfo {title} {Non-darcy flow and its implications to seepage problems},\ }\href@noop {} {\bibfield  {journal} {\bibinfo  {journal} {Journal of the Irrigation and Drainage Division}\ }\textbf {\bibinfo {volume} {103}},\ \bibinfo {pages} {459} (\bibinfo {year} {1977})}\BibitemShut {NoStop}%
\bibitem [{\citenamefont {Wang}\ \emph {et~al.}(2019)\citenamefont {Wang}, \citenamefont {Li}, \citenamefont {Zhao}, \citenamefont {Chen},\ and\ \citenamefont {Xu}}]{wang2019experimental}%
  \BibitemOpen
  \bibfield  {author} {\bibinfo {author} {\bibfnamefont {L.}~\bibnamefont {Wang}}, \bibinfo {author} {\bibfnamefont {Y.}~\bibnamefont {Li}}, \bibinfo {author} {\bibfnamefont {G.}~\bibnamefont {Zhao}}, \bibinfo {author} {\bibfnamefont {N.}~\bibnamefont {Chen}},\ and\ \bibinfo {author} {\bibfnamefont {Y.}~\bibnamefont {Xu}},\ }\bibfield  {title} {\bibinfo {title} {Experimental investigation of flow characteristics in porous media at low reynolds numbers ($re\rightarrow0$) under different constant hydraulic heads},\ }\href {https://doi.org/10.3390/w11112317} {\bibfield  {journal} {\bibinfo  {journal} {Water}\ }\textbf {\bibinfo {volume} {11}},\ \bibinfo {pages} {2317} (\bibinfo {year} {2019})}\BibitemShut {NoStop}%
\bibitem [{\citenamefont {Naqvi}\ \emph {et~al.}(2026)\citenamefont {Naqvi}, \citenamefont {{\'S}nie{\.z}ek}, \citenamefont {Strzelczyk}, \citenamefont {M{\k{a}}drala},\ and\ \citenamefont {Matyka}}]{Naqvi26}%
  \BibitemOpen
  \bibfield  {author} {\bibinfo {author} {\bibfnamefont {S.~B.}\ \bibnamefont {Naqvi}}, \bibinfo {author} {\bibfnamefont {D.}~\bibnamefont {{\'S}nie{\.z}ek}}, \bibinfo {author} {\bibfnamefont {D.}~\bibnamefont {Strzelczyk}}, \bibinfo {author} {\bibfnamefont {M.}~\bibnamefont {M{\k{a}}drala}},\ and\ \bibinfo {author} {\bibfnamefont {M.}~\bibnamefont {Matyka}},\ }\bibfield  {title} {\bibinfo {title} {Inertial effects on fluid flow through natural porous media},\ }\href {https://doi.org/10.1029/2025WR041274} {\bibfield  {journal} {\bibinfo  {journal} {Water Resources Research}\ }\textbf {\bibinfo {volume} {62}},\ \bibinfo {pages} {e2025WR041274} (\bibinfo {year} {2026})}\BibitemShut {NoStop}%
\bibitem [{\citenamefont {Koponen}\ \emph {et~al.}(1997)\citenamefont {Koponen}, \citenamefont {Kataja},\ and\ \citenamefont {Timonen}}]{Koponen97}%
  \BibitemOpen
  \bibfield  {author} {\bibinfo {author} {\bibfnamefont {A.}~\bibnamefont {Koponen}}, \bibinfo {author} {\bibfnamefont {M.}~\bibnamefont {Kataja}},\ and\ \bibinfo {author} {\bibfnamefont {J.}~\bibnamefont {Timonen}},\ }\bibfield  {title} {\bibinfo {title} {Permeability and effective porosity of porous media},\ }\href@noop {} {\bibfield  {journal} {\bibinfo  {journal} {Physical Review E}\ }\textbf {\bibinfo {volume} {56}},\ \bibinfo {pages} {3319} (\bibinfo {year} {1997})}\BibitemShut {NoStop}%
\bibitem [{\citenamefont {Duda}\ \emph {et~al.}(2011)\citenamefont {Duda}, \citenamefont {Koza},\ and\ \citenamefont {Matyka}}]{Duda11}%
  \BibitemOpen
  \bibfield  {author} {\bibinfo {author} {\bibfnamefont {A.}~\bibnamefont {Duda}}, \bibinfo {author} {\bibfnamefont {Z.}~\bibnamefont {Koza}},\ and\ \bibinfo {author} {\bibfnamefont {M.}~\bibnamefont {Matyka}},\ }\bibfield  {title} {\bibinfo {title} {Hydraulic tortuosity in arbitrary porous media flow},\ }\href@noop {} {\bibfield  {journal} {\bibinfo  {journal} {Physical Review E}\ }\textbf {\bibinfo {volume} {84}},\ \bibinfo {pages} {036319} (\bibinfo {year} {2011})}\BibitemShut {NoStop}%
\bibitem [{\citenamefont {Andrade}\ \emph {et~al.}(1997)\citenamefont {Andrade}, \citenamefont {Almeida}, \citenamefont {Mendes~Filho}, \citenamefont {Havlin}, \citenamefont {Suki},\ and\ \citenamefont {Stanley}}]{andrade1997fluid}%
  \BibitemOpen
  \bibfield  {author} {\bibinfo {author} {\bibfnamefont {J.~S.~J.}\ \bibnamefont {Andrade}}, \bibinfo {author} {\bibfnamefont {M.~P.}\ \bibnamefont {Almeida}}, \bibinfo {author} {\bibfnamefont {J.}~\bibnamefont {Mendes~Filho}}, \bibinfo {author} {\bibfnamefont {S.}~\bibnamefont {Havlin}}, \bibinfo {author} {\bibfnamefont {B.}~\bibnamefont {Suki}},\ and\ \bibinfo {author} {\bibfnamefont {H.~E.}\ \bibnamefont {Stanley}},\ }\bibfield  {title} {\bibinfo {title} {Fluid flow through porous media: The role of stagnant zones},\ }\href {https://doi.org/10.1103/PhysRevLett.79.3901} {\bibfield  {journal} {\bibinfo  {journal} {Physical Review Letters}\ }\textbf {\bibinfo {volume} {79}},\ \bibinfo {pages} {3901} (\bibinfo {year} {1997})}\BibitemShut {NoStop}%
\bibitem [{\citenamefont {Madadi}\ and\ \citenamefont {Sahimi}(2003)}]{madadi2003rough}%
  \BibitemOpen
  \bibfield  {author} {\bibinfo {author} {\bibfnamefont {M.}~\bibnamefont {Madadi}}\ and\ \bibinfo {author} {\bibfnamefont {M.}~\bibnamefont {Sahimi}},\ }\bibfield  {title} {\bibinfo {title} {Lattice boltzmann simulation of fluid flow in fracture networks with rough, self-affine surfaces},\ }\href {https://doi.org/10.1103/PhysRevE.67.026309} {\bibfield  {journal} {\bibinfo  {journal} {Physical Review E}\ }\textbf {\bibinfo {volume} {67}},\ \bibinfo {pages} {026309} (\bibinfo {year} {2003})}\BibitemShut {NoStop}%
\bibitem [{\citenamefont {Madadi}\ \emph {et~al.}(2003)\citenamefont {Madadi}, \citenamefont {Van~Siclen},\ and\ \citenamefont {Sahimi}}]{madadi2003fractures}%
  \BibitemOpen
  \bibfield  {author} {\bibinfo {author} {\bibfnamefont {M.}~\bibnamefont {Madadi}}, \bibinfo {author} {\bibfnamefont {C.~D.}\ \bibnamefont {Van~Siclen}},\ and\ \bibinfo {author} {\bibfnamefont {M.}~\bibnamefont {Sahimi}},\ }\bibfield  {title} {\bibinfo {title} {Fluid flow and conduction in two-dimensional fractures with rough, self-affine surfaces: A comparative study},\ }\href {https://doi.org/10.1029/2002JB002332} {\bibfield  {journal} {\bibinfo  {journal} {Journal of Geophysical Research: Solid Earth}\ }\textbf {\bibinfo {volume} {108}},\ \bibinfo {pages} {2396} (\bibinfo {year} {2003})}\BibitemShut {NoStop}%
\bibitem [{\citenamefont {Adrover}(2011)}]{adrover2011koch}%
  \BibitemOpen
  \bibfield  {author} {\bibinfo {author} {\bibfnamefont {A.}~\bibnamefont {Adrover}},\ }\bibfield  {title} {\bibinfo {title} {Convection-dominated dispersion in channels with fractal cross-section},\ }\href {https://doi.org/10.1063/1.3526759} {\bibfield  {journal} {\bibinfo  {journal} {Physics of Fluids}\ }\textbf {\bibinfo {volume} {23}},\ \bibinfo {pages} {013603} (\bibinfo {year} {2011})}\BibitemShut {NoStop}%
\bibitem [{\citenamefont {Jasak}\ \emph {et~al.}(2007)\citenamefont {Jasak}, \citenamefont {Jemcov}, \citenamefont {Tukovic} \emph {et~al.}}]{jasak2007openfoam}%
  \BibitemOpen
  \bibfield  {author} {\bibinfo {author} {\bibfnamefont {H.}~\bibnamefont {Jasak}}, \bibinfo {author} {\bibfnamefont {A.}~\bibnamefont {Jemcov}}, \bibinfo {author} {\bibfnamefont {Z.}~\bibnamefont {Tukovic}}, \emph {et~al.},\ }\bibfield  {title} {\bibinfo {title} {{OpenFOAM}: A c++ library for complex physics simulations},\ }in\ \href@noop {} {\emph {\bibinfo {booktitle} {International Workshop on Coupled Methods in Numerical Dynamics}}},\ Vol.\ \bibinfo {volume} {1000}\ (\bibinfo {year} {2007})\ pp.\ \bibinfo {pages} {1--20}\BibitemShut {NoStop}%
\bibitem [{\citenamefont {Patankar}(1980)}]{patankar1980numerical}%
  \BibitemOpen
  \bibfield  {author} {\bibinfo {author} {\bibfnamefont {S.~V.}\ \bibnamefont {Patankar}},\ }\href@noop {} {\emph {\bibinfo {title} {Numerical Heat Transfer and Fluid Flow}}}\ (\bibinfo  {publisher} {Hemisphere Publishing Corporation},\ \bibinfo {address} {Washington, DC},\ \bibinfo {year} {1980})\BibitemShut {NoStop}%
\bibitem [{\citenamefont {Jureti{\'c}}(2015)}]{cfmesh2015}%
  \BibitemOpen
  \bibfield  {author} {\bibinfo {author} {\bibfnamefont {F.}~\bibnamefont {Jureti{\'c}}},\ }\href {https://cfmesh.com/documentation/} {\emph {\bibinfo {title} {cfMesh User Guide}}},\ \bibinfo {organization} {Creative Fields},\ \bibinfo {address} {Zagreb} (\bibinfo {year} {2015})\BibitemShut {NoStop}%
\bibitem [{\citenamefont {Strzelczyk}\ \emph {et~al.}(2025)\citenamefont {Strzelczyk}, \citenamefont {Rot}, \citenamefont {Kosec},\ and\ \citenamefont {Matyka}}]{Strzelczyk2025}%
  \BibitemOpen
  \bibfield  {author} {\bibinfo {author} {\bibfnamefont {D.}~\bibnamefont {Strzelczyk}}, \bibinfo {author} {\bibfnamefont {M.}~\bibnamefont {Rot}}, \bibinfo {author} {\bibfnamefont {G.}~\bibnamefont {Kosec}},\ and\ \bibinfo {author} {\bibfnamefont {M.}~\bibnamefont {Matyka}},\ }\bibfield  {title} {\bibinfo {title} {Cross-validation of meshless navier--stokes solvers in porous media flows},\ }\href {https://doi.org/10.1038/s41598-025-05272-x} {\bibfield  {journal} {\bibinfo  {journal} {Scientific Reports}\ }\textbf {\bibinfo {volume} {15}},\ \bibinfo {pages} {33837} (\bibinfo {year} {2025})}\BibitemShut {NoStop}%
\bibitem [{\citenamefont {Goursat}(1898)}]{goursat1898}%
  \BibitemOpen
  \bibfield  {author} {\bibinfo {author} {\bibfnamefont {E.}~\bibnamefont {Goursat}},\ }\bibfield  {title} {\bibinfo {title} {Sur l'{\'e}quation $\delta\delta u = 0$},\ }\href@noop {} {\bibfield  {journal} {\bibinfo  {journal} {Bulletin de la Soci{\'e}t{\'e} Math{\'e}matique de France}\ }\textbf {\bibinfo {volume} {26}},\ \bibinfo {pages} {236} (\bibinfo {year} {1898})}\BibitemShut {NoStop}%
\bibitem [{\citenamefont {Moffatt}(1964)}]{moffatt1964viscous}%
  \BibitemOpen
  \bibfield  {author} {\bibinfo {author} {\bibfnamefont {H.~K.}\ \bibnamefont {Moffatt}},\ }\bibfield  {title} {\bibinfo {title} {Viscous and resistive eddies near a sharp corner},\ }\href {https://doi.org/10.1017/S0022112064000015} {\bibfield  {journal} {\bibinfo  {journal} {Journal of Fluid Mechanics}\ }\textbf {\bibinfo {volume} {18}},\ \bibinfo {pages} {1} (\bibinfo {year} {1964})}\BibitemShut {NoStop}%
\bibitem [{\citenamefont {Vekua}(1967)}]{vekua1967}%
  \BibitemOpen
  \bibfield  {author} {\bibinfo {author} {\bibfnamefont {I.~N.}\ \bibnamefont {Vekua}},\ }\href@noop {} {\emph {\bibinfo {title} {New Methods for Solving Elliptic Equations}}}\ (\bibinfo  {publisher} {North-Holland},\ \bibinfo {address} {Amsterdam},\ \bibinfo {year} {1967})\BibitemShut {NoStop}%
\bibitem [{\citenamefont {Brubeck}\ and\ \citenamefont {Trefethen}(2022)}]{brubeck2022lightning}%
  \BibitemOpen
  \bibfield  {author} {\bibinfo {author} {\bibfnamefont {P.~D.}\ \bibnamefont {Brubeck}}\ and\ \bibinfo {author} {\bibfnamefont {L.~N.}\ \bibnamefont {Trefethen}},\ }\bibfield  {title} {\bibinfo {title} {Lightning {Stokes} solver},\ }\href {https://doi.org/10.1137/21M1408579} {\bibfield  {journal} {\bibinfo  {journal} {SIAM Journal on Scientific Computing}\ }\textbf {\bibinfo {volume} {44}},\ \bibinfo {pages} {A1205} (\bibinfo {year} {2022})}\BibitemShut {NoStop}%
\bibitem [{\citenamefont {Fox}\ \emph {et~al.}(1967)\citenamefont {Fox}, \citenamefont {Henrici},\ and\ \citenamefont {Moler}}]{fox1967approximations}%
  \BibitemOpen
  \bibfield  {author} {\bibinfo {author} {\bibfnamefont {L.}~\bibnamefont {Fox}}, \bibinfo {author} {\bibfnamefont {P.}~\bibnamefont {Henrici}},\ and\ \bibinfo {author} {\bibfnamefont {C.}~\bibnamefont {Moler}},\ }\bibfield  {title} {\bibinfo {title} {Approximations and bounds for eigenvalues of elliptic operators},\ }\href {https://doi.org/10.1137/0704008} {\bibfield  {journal} {\bibinfo  {journal} {SIAM Journal on Numerical Analysis}\ }\textbf {\bibinfo {volume} {4}},\ \bibinfo {pages} {89} (\bibinfo {year} {1967})}\BibitemShut {NoStop}%
\end{thebibliography}
\end{document}